\documentclass[10pt,journal,compsoc]{IEEEtran}
\ifCLASSOPTIONcompsoc
  \usepackage[nocompress]{cite}
\else
  \usepackage{cite}
\fi

\PassOptionsToPackage{hyphens}{url}
\usepackage[bookmarks=false,hidelinks]{hyperref}
\usepackage[table]{xcolor}
\usepackage{subfigure}
\usepackage{tikz}
\usepackage{pgfplots}
\pgfplotsset{compat=newest}
\usetikzlibrary{positioning,calc}
\usetikzlibrary{shapes,arrows}
\usepgfplotslibrary{groupplots}
\usepackage{booktabs}
\usepackage{multirow}
\usepackage{amsmath,amssymb}
\usepackage{fdsymbol}
\usepackage{makecell}
\usepgfplotslibrary{statistics}

\usepackage{graphicx}

\def\addlegendimage{\csname pgfplots@addlegendimage\endcsname}

\newcommand{\setvariable}[2]{
	\let#1\relax
	\newcommand{#1}{#2}
}
\usepackage{xcolor}
\usepackage{colorbrewer}
\usepackage{balance}

\IEEEoverridecommandlockouts
\IEEEpubid{\makebox[\columnwidth]{\copyright2021 IEEE Transactions on Biometrics, Behavior, and Identity Science (T-BIOM)\hfill} \hspace{\columnsep}\makebox[\columnwidth]{ }}

\begin{document}

\title{ASVspoof 2019: spoofing countermeasures \\ for the detection of synthesized, converted \\ and replayed speech}
\author{Andreas~Nautsch,~\IEEEmembership{Member,~IEEE,} Xin~Wang,~\IEEEmembership{Member,~IEEE,} Nicholas~Evans,~\IEEEmembership{Member,~IEEE,} Tomi~Kinnunen,~\IEEEmembership{Member,~IEEE,} Ville~Vestman,
Massimiliano~Todisco,~\IEEEmembership{Member,~IEEE,} H\'ector~Delgado, Md~Sahidullah,~\IEEEmembership{Member,~IEEE,} Junichi~Yamagishi,~\IEEEmembership{Senior~Member,~IEEE,} and~Kong~Aik~Lee,~\IEEEmembership{Senior~Member,~IEEE}%
\thanks{Manuscript received MMMM DD, YYYY; revised MMMM DD, YYYY; accepted MMMM, DD YYYY. Date of publication MMMM DD, YYYY; date of current version MMMM DD, YYYY. This  work was supported  by a number of projects and funding sources:  VoicePersonae, supported by the French Agence Nationale de la Recherche~(ANR) and the Japan Science and Technology Agency (JST) with grant No.\ JPMJCR18A6; RESPECT, supported by the ANR; the NOTCH project (no.\ 309629), supported by the Academy of Finland; Region Grand Est, France.  (Corresponding author: Andreas Nautsch.)}%
\thanks{A. Nautsch, N. Evans and M. Todisco are with EURECOM, Campus SophiaTech, 450 Route des Chappes, 06410 Biot, France. E-mail: \{nautsch,evans,todisco\}@eurecom.fr}
\thanks{X. Wang and J. Yamagishi are with National Institute of Informatics, 2-1-2 Hitotsubashi, Chiyoda-ku, Tokyo, Japan. E-mail: \{wangxin,jyamagis\}@nii.ac.jp}
\thanks{T. Kinnunen and V. Vestman are with University of Eastern Finland, Joensuu campus, L\"{a}nsikatu 15, FI-80110 Joensuu, Finland. E-mail: \{tkinnu,ville.vestman\}@uef.fi}
\thanks{H. Delgado is with Nuance Communications, C/ Gran V\'ia 39, 28013 Madrid, Spain. E-mail: hector.delgado@nuance.com.}
\thanks{Md Sahidullah is with Universit\'{e} de Lorraine, CNRS, Inria, LORIA, F-54000, Nancy, France. E-mail: md.sahidullah@inria.fr}
\thanks{K. A. Lee is with Institute for Infocomm Research, A$^{\star}$STAR, 1 Fusionopolis Way, Singapore 138632. E-mail: lee\_kong\_aik@i2r.a-star.edu.sg}
}

\markboth{IEEE Transactions on Biometrics, Behavior, and Identity Science (T-BIOM),~Vol.~VV, No.~NN, MMMM~YYYY}%
{Nautsch \MakeLowercase{\textit{et al.}}: ASVspoof 2019: spoofing countermeasures for the detection of synthesized, converted and replayed speech}

\IEEEtitleabstractindextext{%
\begin{abstract}

The ASVspoof initiative was conceived to spearhead research in anti-spoofing for automatic speaker verification (ASV).
This paper describes the third in a series of bi-annual challenges: ASVspoof~2019.  
With the challenge database and protocols being described elsewhere, the focus of this paper is on results and the top performing single and ensemble system submissions from 62 teams, all of which out-perform the two baseline systems, often by a substantial margin. 
Deeper analyses shows that performance is dominated by specific conditions involving either specific spoofing attacks or specific acoustic environments.
While fusion is shown to be particularly effective for the logical access scenario
involving speech synthesis and voice conversion attacks,
participants largely struggled to apply fusion successfully for the physical access scenario involving simulated replay attacks.
This is likely the result of a lack of system complementarity, while oracle fusion experiments show clear potential to improve performance.
Furthermore, while results for simulated data are promising, experiments with real replay data show a substantial gap, most likely due to the presence of additive noise in the latter.
This finding, among others, leads to a number of ideas for further research and directions for future editions of the ASVspoof challenge.
\end{abstract}
}

\IEEEdisplaynontitleabstractindextext
\IEEEpeerreviewmaketitle
\maketitle

\IEEEraisesectionheading{\section{Introduction}
\label{S:1}}

\IEEEPARstart{I}{t} is well known that automatic speaker verification (ASV) systems are vulnerable to being manipulated by spoofing, also known as presentation attacks~\cite{Sahidullah-book-2019}.  Spoofing attacks can enable a fraudster to gain illegitimate access to resources, services or devices protected by ASV technology.  The threat from spoofing can be substantial and unacceptable.  Following the first special session on anti-spoofing held in 2013~\cite{evans2013spoofing}, the effort to develop spoofing countermeasures, auxiliary systems which aim to protect ASV technology by automatically detecting and deflecting spoofing attacks, has been spearheaded by the ASVspoof initiative\footnote{\url{https://www.asvspoof.org}}.  

\begin{figure*}[!t]
	\centering
	\includegraphics[width=0.75\linewidth]{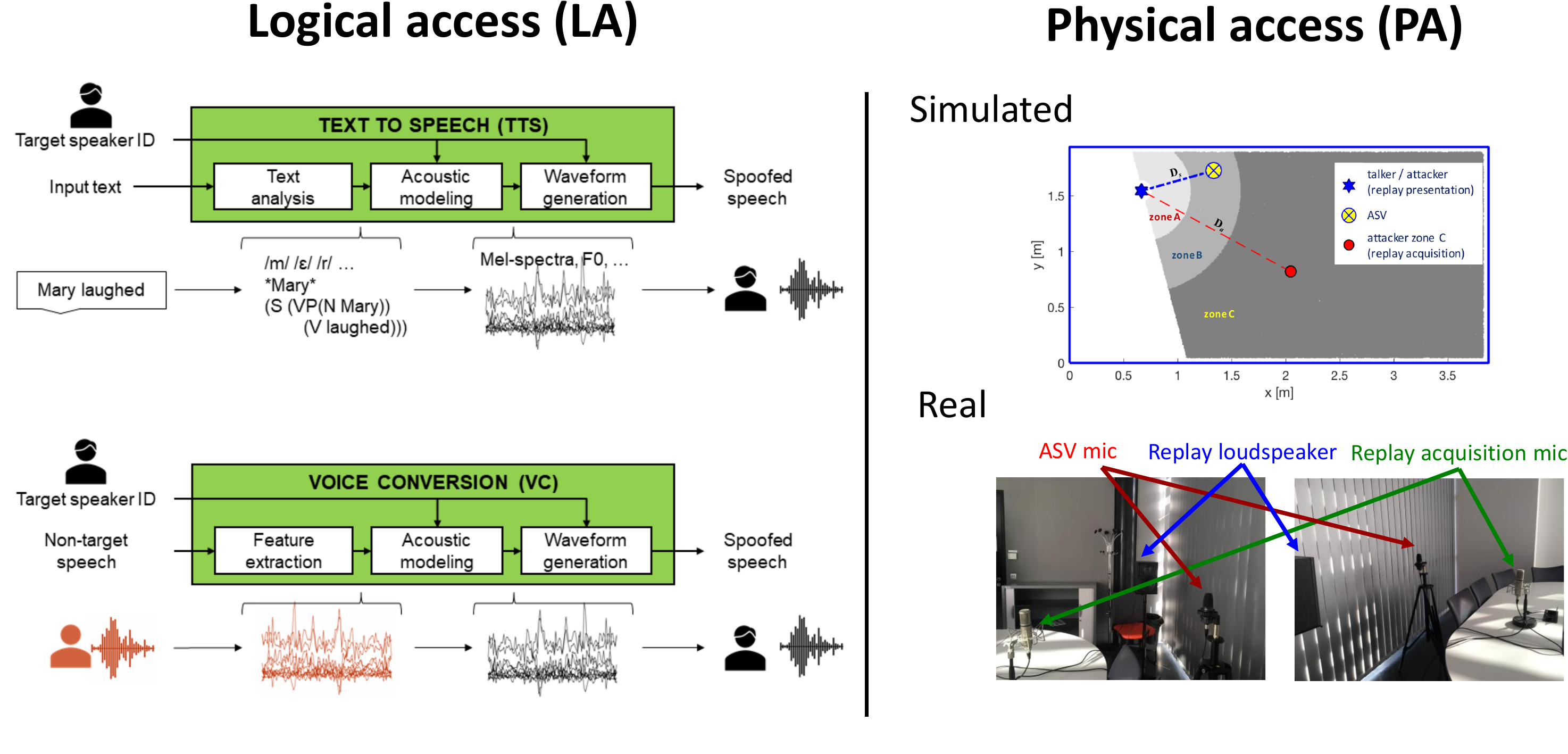}
	\caption{The ASVspoof 2019 challenge featured four different types of spoofed audio data. 
	The LA scenario contains text-to-speech and voice conversion attacks. In the PA scenario attackers acquire a recording of the target speaker which is then replayed to the ASV system. Both \emph{simulated} and \emph{real} replay attacks are considered. The former refers to simulated acoustic environments/rooms with specified dimensions and controllable reverberation whereas the latter contains actual replay recordings collected at three different sites. Real replay attacks were included in the test set (but excluded from challenge ranking). 
	This paper describes challenge results for all four setups illustrated.}
	\label{fig:physical-logicalScenario}
\end{figure*}

ASVspoof 2019~\cite{Todisco2019}, the most recent of three editions and the focus in this paper, was the first to include all three major forms of spoofing attacks involving speech synthesis, voice conversion and replay, in separate logical and physical access scenarios. It also brought several advances with respect to previous editions. First, ASVspoof 2019 aimed to explore whether advances in speech synthesis and voice conversion technologies pose a greater threat to ASV reliability; the latest of these techniques, \emph{e.g.}\ neural network-based waveform modelling techniques, can produce synthetic and converted speech that is perceptually indistinguishable from bona fide speech.  Second, the 2019 edition explored replay attacks using a far more controlled evaluation setup in the form of simulated replay attacks and carefully controlled acoustic conditions.
Third, the database is substantially larger than the previous ASVspoof databases and considerably more diverse in terms of attack algorithms.  With a comprehensive description of the database available in a published companion paper~\cite{wang2020asvspoof}, only a brief description is provided in the current article.  The focus here is instead upon challenge results and findings. 

Whereas previous editions of ASVspoof utilised the equal error rate (EER) metric to judge performance, the 2019 edition shifted to the ASV-centric tandem detection cost function (t-DCF) metric~\cite{Kinnunen2018-tDCF,Kinnunen2020_tDCF}.  While the latter reflects the impact of both spoofing and countermeasures upon ASV performance, participation still calls only for the development and optimisation of {\it countermeasures}.  Strong performance depends upon generalisation, namely countermeasures that perform well in the face of spoofing attacks not seen in training or development data.  

Two different baseline systems were provided for the 2019 edition. With data, protocols and metrics being different to those of previous editions, progress is judged in terms of performance relative to the two baseline systems which provide some level of continuity or consistency with previous challenge editions.  The article describes the top five single and fused systems for both challenge scenarios, provides insights into the most successful countermeasure (CM) techniques and assesses their impact upon ASV reliability.  Finally, the article outlines priorities for the ASVspoof initiative looking to the future, including ideas for the next edition -- ASVspoof 2021.

\section{Challenge outline}

This section describes the logical and physical access ASVspoof 2019 challenge scenarios (see Fig.~\ref{fig:physical-logicalScenario}), the challenge rules, the new t-DCF metric, and baseline CM and ASV systems. Since it is described elsewhere~\cite{wang2020asvspoof}, the ASVspoof 2019 database is not described here.  
Briefly, it is sourced from the \emph{Voice Cloning Toolkit} (VCTK) corpus~\cite{vctk}, a multi-speaker, native-English speech database of read sentences recorded in a hemi-anechoic chamber.

\subsection{Logical access}

Logical access (LA) control implies a scenario in which a remote user seeks access to a system or service protected by ASV.  
An example is a telephone banking service to which attackers may connect 
and then send synthetic or converted voice signals directly to the ASV system while bypassing the microphone, \emph{i.e.}\ by injecting audio into the communication channel post sensor.

The LA subset of the ASVspoof 2019 database was created using a diverse array of 17 text-to-speech~(TTS), voice conversion~(VC) and hybrid systems. Waveform generation methods vary from waveform concatenation to neural network-based waveform  modelling techniques including WaveNet~\cite{oord2016wavenet}. Acoustic models also vary from Gaussian mixture models to advanced sequence-to-sequence neural networks. Some are constructed using popular open-source toolkits while others are selected based on their superior evaluation results reported in the Voice Conversion Challenge~\cite{LorenzoTrueba2018-VCC} or other literature. 
Six of these systems are designated as known spoofing algorithms/attacks, with the other 11 being designated as unknown spoofing attacks. Among the 6 known attacks there are 2 VC systems and 4 TTS systems. The 11 unknown attacks comprise 2 VC, 6 TTS and 3 hybrid TTS-VC systems for which VC systems are fed with synthetic speech.  Known attacks are used to generate training and development data.  Unknown attacks and  
two of the known attacks
are used to generate evaluation data.  
Attacks are referred to by attack identifiers (AIDs): A1 -- A19.
Full details of the LA setup, attack groups and analysis are provided in~\cite{wang2020asvspoof}.

\subsection{Physical access}
\label{sec:pa_description}

In the physical access (PA) scenario, spoofing 
attacks are presented to a fixed microphone which is placed in an environment in which sounds propagate and are reflected from obstacles such as floors and walls.  Spoofing attacks in this scenario are referred to as replay attacks and match the ISO definition of \emph{presentation attacks}~\cite{ISO-IEC-30107-1-PAD-Framework-160115}.
The PA scenario, is based upon \emph{simulated} and carefully controlled acoustic and replay configurations~\cite{campbell05,vincent08,novak2015synchronized}.
The approach used to simulate room acoustics under varying source/receiver positions is inspired from the approach reported in~\cite{janicki2016assessment} and based upon an image-source method~\cite{Allen79}. Acoustic simulations are performed using Roomsimove\footnote{\url{http://homepages.loria.fr/evincent/software/Roomsimove_1.4.zip}}, while the replay device effects are simulated using the generalised polynomial Hammerstein model and the Synchronized Swept Sine tool\footnote{\url{https://ant-novak.com/pages/sss/}}.

The ASV system is used within a noise-free acoustic environment defined by: the room size~$S$; the $T60$ reverberation time; the talker-to-ASV\footnote{We refer to a talker instead of speaker in order to avoid confusion with the loudspeaker.} distance $D_\text{s}$. Each parameter is categorised into three different intervals. The room size~$S$ is categorised into: (a)~small rooms of size 2-5~$\textup{m}^2$; (b)~medium rooms of size 5-10~$\textup{m}^2$; (c)~large rooms of size 10-20~$\textup{m}^2$.  The $T60$ reverberation time is categorised into: (a)~low 50-200~$\textup{ms}$; (b)~medium 200-600~$\textup{ms}$; (c)~large 600-1000~$\textup{ms}$.  The talker-to-ASV distance~$D_\text{s}$ is categorised into: (a)~low 10-50~$\textup{cm}$; (b)~medium 50-100~$\textup{cm}$; (c)~large 100-150~$\textup{cm}$. 
This results in 27 acoustic configurations denoted by environment identifiers~(EIDs) (\textit{aaa}, \textit{aab}, ..., \textit{ccc}).

A replay spoofing attack is mounted through the making of a surreptitious recording of a bona fide access attempt and then the presentation of this recording to the ASV microphone.  Attackers acquire recordings of bona fide access attempts when positioned at an attacker-to-talker distance~$D_\text{a}$ from the talker whereas the presentation of recording is made at the talker-to-ASV 
distance~$D_\text{s}$ using a playback device of quality~$Q$.
$D_\text{a}$ is categorised into three different intervals: (A)~low 10-50~$\textup{cm}$; (B)~medium 50-100~$\textup{cm}$; (C)~large 100-150~$\textup{cm}$.  $Q$~is categorised into three quality groups: (A)~perfect quality, \emph{i.e.}\ a Dirac impulse response; (B)~high quality; (C)~low quality. Their combination results in 9 attack configurations denoted by attack identifiers~(AIDs) (\textit{AA}, \textit{AB}, ..., \textit{CC}). 
Full details of the PA setup are also provided in~\cite{wang2020asvspoof}.

\subsection{Rules}

\begin{table}[!t]
\center
\caption{\label{tab:allowed-submissions} 
Submission categories for the ASVspoof 2019 challenge for both LA and PA scenarios and primary, single and contrastive submission.  Only results for single and primary systems are discussed in this paper.
} 
\begin{tabular}{|c|c|c|c|}
    \hline
                    \multicolumn{4}{|c|}{LOGICAL ACCESS (LA) sub-challenge}\\
                    & ASV            & \multicolumn{2}{c|}{CM scores}\\
    Submission      & scores        & Dev       & Eval\\
    \hline \hline     
    Single system   & ---   & \textbf{Required}     & \textbf{Required}\\
    Primary         & ---   & \textbf{Required}     & \textbf{Required}\\
    \textcolor{gray}{Contrastive1}    & \textcolor{gray}{---}   & \textcolor{gray}{Optional}              &   \textcolor{gray}{Optional}\\
    \textcolor{gray}{Contrastive2}    & \textcolor{gray}{---}   & \textcolor{gray}{Optional}              &   \textcolor{gray}{Optional}\\
    \hline
    \hline
                     \multicolumn{4}{|c|}{PHYSICAL ACCESS (PA) sub-challenge}\\
                    & ASV            & \multicolumn{2}{c|}{CM scores}\\
    Submission      & scores        & Dev       & Eval\\
    \hline \hline     
    Single system   & ---   & \textbf{Required}     & \textbf{Required}\\
    Primary         & ---   & \textbf{Required}     & \textbf{Required}\\
    \textcolor{gray}{Contrastive1}    & \textcolor{gray}{---}   & \textcolor{gray}{Optional}              &   \textcolor{gray}{Optional}\\
    \textcolor{gray}{Contrastive2}    & \textcolor{gray}{---}   & \textcolor{gray}{Optional}              &   \textcolor{gray}{Optional}\\
    \hline
\end{tabular}
\end{table}

The submission categories for ASVspoof 2019 are illustrated in Table~\ref{tab:allowed-submissions}.
Participants were permitted to submit up to 4 different score sets (or 8, counting sets for development and evaluation separately) for the LA scenario and an additional 4 for the PA scenario, with the use of different systems being permitted for each.  Two of these score sets are required and include \emph{primary} and \emph{single} system scores. 
Score submissions were required for both the development and evaluation subsets defined in the ASVspoof 2019 protocols.
Scores for corresponding development and evaluation subsets were required to be derived using identical CM systems without any adaptation.
Ensemble classifiers consisting of multiple sub-systems whose output scores are combined were permitted for primary systems only.
Single system scores were required to be one of the sub-systems in the ensemble (normally the single, best performing).
While participants were permitted to submit scores for an additional two \emph{contrastive} systems,
only results for single and primary systems are presented in this paper.

ASV scores used for scoring and ranking were computed by the organisers using separate ASV protocols.   
The use of external data resources was forbidden: all 
systems 
designed by the participants were required to be trained and optimised using \emph{only} the relevant ASVspoof 2019 data and protocols. 
The only exception to this rule is the use of data augmentation, but only then using ASVspoof 2019 training and development data with external, \emph{non-speech} data, \emph{e.g.} impulse responses. Use of LA data for PA experiments and vice versa was also forbidden.

Finally, CM scores produced for any one trial must be obtained using \emph{only} the data in that trial segment. The use of data from any other trial segments was strictly prohibited. Therefore, the use of techniques such as normalization over multiple trial segments and the use of trial data for model adaptation was forbidden. Systems must therefore process trial lists segment-by-segment independently without access to past or future trial segments.

\subsection{Metrics}
\label{sec:metrics}

While the parameter-free equal error rate (EER) metric is retained as a secondary metric,
the primary metric is
the \emph{tandem detection cost function} (t-DCF)~\cite{Kinnunen2018-tDCF}, and the specific \textbf{ASV-constrained} variant detailed in~\cite{Kinnunen2020_tDCF}. 
The 
detection threshold (set to the EER operating point) of the ASV system (designed by the organiser) 
is fixed,
whereas the detection threshold of the CM system (designed by participants) is allowed to vary. 
Results are reported in the form of \textbf{minimum normalized} t-DCF values, defined as
    \begin{equation}\label{eq:tDCF-ASV-constrained}
        \text{min t-DCF} = \min_{\tau_\text{cm}}\Bigg\{\frac{ C_0+C_1P_\text{miss}^\text{cm}(\tau_\text{cm})
        +C_2P_\text{fa}^\text{cm}(\tau_\text{cm})}{\text{t-DCF}_\text{default}}\Bigg\},
    \end{equation} 
where $P_\text{miss}(\tau_\text{cm})$ and $P_\text{fa}(\tau_\text{cm})$ are the miss and false alarm rates of the CM at threshold $\tau_\text{cm}$. Coefficients $C_0$, $C_1$ and $C_2$~\cite[Eq. (11)]{Kinnunen2020_tDCF} depend not only on pre-defined target, nontarget and spoofing attack priors and detection costs but \emph{also} on the miss, false alarm and spoof false alarm rates (the ratio of spoofed trials accepted by the ASV to the total number of spoofed trials) of the ASV system. 

The denominator $\text{t-DCF}_\text{default}=C_0+\min\{C_1,C_2\}$ is the cost of an uninformative \emph{default} CM that either accepts or rejects every test utterance. Its inclusion ensures that min t-DCF values are in the range between 0 and 1. A value~0 means that both ASV and CM systems are error-free whereas a value~1 means that the CM cannot improve upon the default system. Another useful reference value in between these two extremes is the case of an error-free CM (but an imperfect ASV), given by $C_0/\text{t-DCF}_\text{default}$. This lower bound is referred to as the \textbf{ASV floor}.     

The above formulation differs slightly from that in the ASVspoof 2019 evaluation plan.  Differences include the absence of sub-system-level detection costs and the inclusion of the ASV floor. 
The numerical scale of the t-DCF values between the formulations differs but the impact upon system rankings is negligible. 
The scoring toolkits\footnote{\url{https://www.asvspoof.org/resources/tDCF_matlab_v2.zip}}$^,$\footnote{\url{https://www.asvspoof.org/resources/tDCF_python_v2.zip}} have been updated to reflect these changes.

One last, relevant detail concerning the t-DCF is how performance across different attack conditions is aggregated. The straightforward way (used for the ranking of ASVspoof 2019 challenge entries as reported in~\cite{Todisco2019}) is to report performance by pooling CM scores across all attack conditions. As an alternative, we further report \textbf{max min t-DCF} across attack conditions in selected cases. Here, `min' refers again to oracle CM calibration while `max' refers to the highest per-condition t-DCF. The `max min' t-DCF, therefore, serves as a reference point for worst-case attacks (see Section~\ref{subsec:worstCase}).

\subsection{Spoofing countermeasures}

Two CM systems were provided to ASVspoof 2019 participants.  Baseline \textbf{B01} uses constant~Q cepstral coefficients (CQCCs)~\cite{TODISCO2017516,Todisco-art-2016} and a bandwidth of 15~Hz to 8~kHz. 
The number of bins per octave is set to 96 and the re-sampling period set to 16. 
Static features of 29 coefficients and the zeroth coefficient are augmented with delta and delta-delta coefficients resulting in 90-dimensional features.

Baseline \textbf{B02} uses linear frequency cepstral coefficients (LFCCs)~\cite{sahidullah2015comparison} and a bandwidth of 30~Hz to 8~kHz.
LFCCs are extracted using a 512-point discrete Fourier transform applied to windows of 20~ms with 50\% overlap. Static features of 19 coefficients and the zeroth coefficient are augmented with delta and delta-delta coefficients resulting in 60-dimensional features.

Both baselines use a Gaussian mixture model (GMM) back-end binary classifier.
Randomly initialised, 512-component models are trained separately using an expectation-maximisation (EM) algorithm and bona fide and spoofed utterances from the ASVspoof 2019 training data.  Scores are log-likelihood ratios given bona fide and spoofed models.  A Matlab package including both baselines is available for download from the ASVspoof website.\footnote{\url{https://www.asvspoof.org}.}

\subsection{ASV system}

The ASV system was used by the organisers to derive the ASV scores used in computing the t-DCF metric. 
It utilizes an
x-vector~\cite{Snyder2018XVectorsRD} embedding extractor network that was pre-trained\footnote{\url{https://kaldi-asr.org/models/m7}} for the \emph{VoxCeleb recipe} of the Kaldi toolkit~\cite{povey2011kaldi}. Training was performed using the speech data collected from 7325 speakers contained within the entire VoxCeleb2 corpus~\cite{chung18voxceleb2} and the development portion of the VoxCeleb1 corpus~\cite{nagrani17voxceleb}. The network extracts 512-dimensional x-vectors which are fed to a probabilistic linear discriminant analysis (PLDA)~\cite{pldaIoffe,prince2007probabilistic} back-end (trained separately for LA and PA scenarios) for ASV scoring. 
PLDA backends were adapted to LA and PA scenarios by using bona fide recordings of CM training data.
ASV scores for the development set were provided to participants so that they could calculate t-DCF values and use these for CM optimisation.  They were not provided for the evaluation set.

\section{Logical access scenario}
\label{sec:la}

This section describes submissions to ASVspoof 2019 for the LA scenario and results.  Single system submissions are described first, followed by primary system submissions,  
presenting only the top-5 performing of 48 LA system submissions in each case. 

\subsection{Single systems}
\label{sec:la:single}

\begin{figure}[!t]
	\centering
	\includegraphics[trim=70 80 70 10,clip,width=1.0\linewidth]{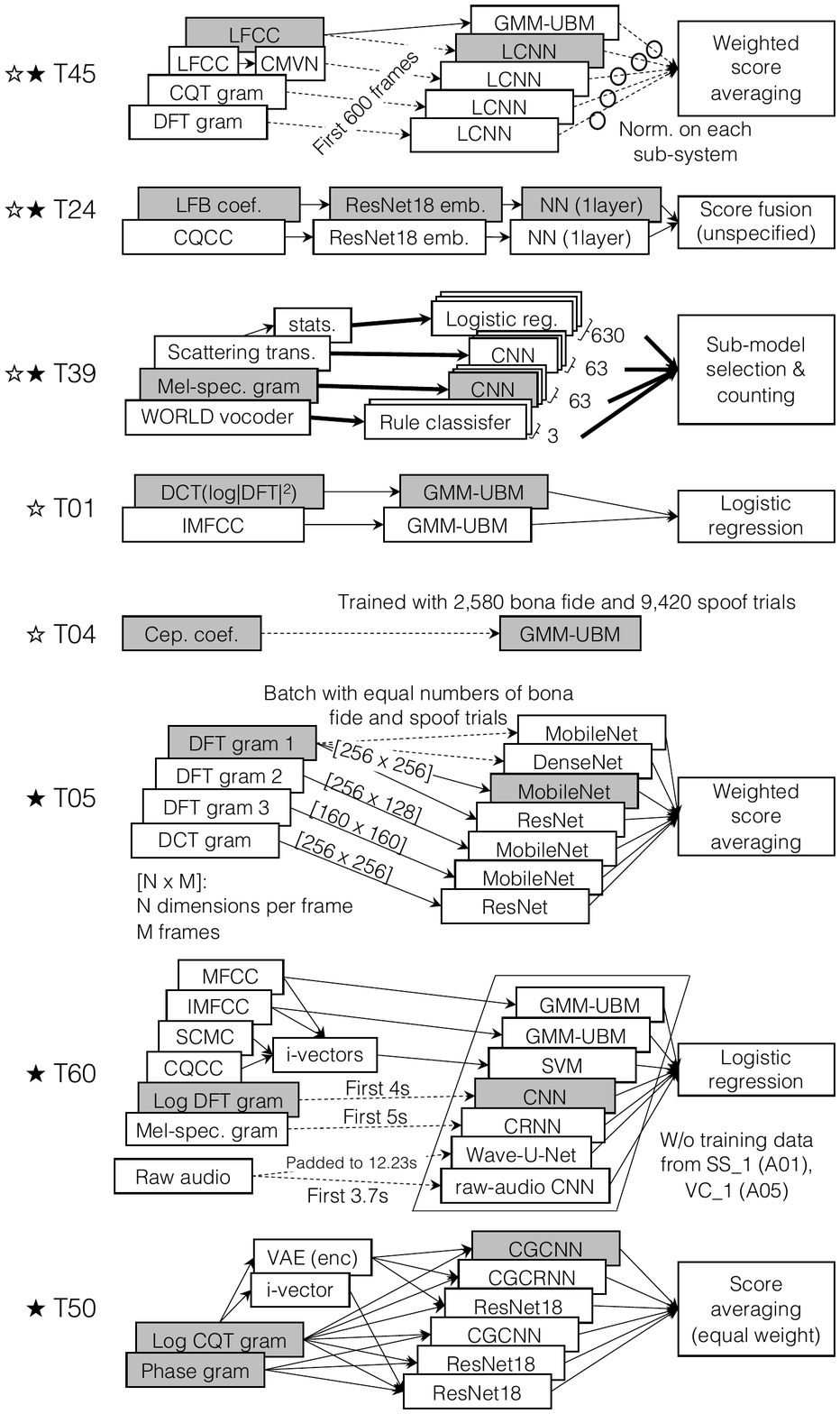}
	\caption{
	Illustration of top single (grey blocks) and primary system submissions for the LA scenario. $\largewhitestar$ and $\largeblackstar$ denote top-5 single and top-5 primary systems, respectively. 
	} 
	\label{fig:LA:illustration}
\end{figure}

The architectures of the top-5 single systems are illustrated in Fig.~\ref{fig:LA:illustration} (grey blocks). Systems are labelled (left) by the anonymised team identifier (TID)~\cite{Todisco2019}.  A short description of each follows: 

\noindent\textbf{T45~\cite{lavrentyeva2019stc}:} A light CNN (LCNN) which operates upon LFCC features extracted from the first 600 frames and with the same frontend configuration as the B2 baseline CM~\cite{sahidullah2015comparison}.  The LCNN uses an angular-margin-based softmax loss (A-softmax)~\cite{liu2017sphereface}, batch normalization~\cite{ioffe2015batch} after max pooling and a normal Kaiming initialization~\cite{he2015delving}. 
 
\noindent\textbf{T24~\cite{Chen2020Odyssey}:} A ResNet classifier which operates upon linear filterbank (LFB) coefficients (no cepstral analysis).
The system extracts embeddings using a modified ResNet-18~\cite{he2016deep} architecture in which the kernel size of the input layer is 3$\times$3 and the stride size of all layers is 1$\times$2. Global mean and standard deviation pooling~\cite{lin2013network} are applied after the last convolutional layer and pooled features go through two fully-connected layers with batch normalization. The output is length-normalized and classified using a single-layer neural network.

\noindent\textbf{T39:} A CNN classifier with Mel-spectrogram features. 
The CNN uses multiple blocks of 1D convolution with ReLU activation, dropout, and subsampling with strided convolution. The CNN output is used as the score without binary classification. 
 
\noindent\textbf{T01:} A GMM-UBM classifier with 60-dimensional STFT cepstral coefficients, including static, delta, and delta-delta components.  The GMM-UBM uses the same configuration as the two baseline CMs.

\noindent\textbf{T04:} Cepstral features with a GMM-UBM classifier.  Due to hardware limitations, models are trained using the full set of bona fide utterances but a random selection of only 9,420 spoofed utterances.

\begin{table}[!t]
    \centering
    \caption{A comparison of top-5 (a) single and (b) primary systems for the LA scenario.  Single systems B01 and B02 are the two baselines, whereas \emph{Perfect} refers to the perfect CM (ASV floor for min t-DCF and EER of 0\%).  Systems are labelled by participating team identifiers (TIDs). 
    Results are presented in terms of the minimum t-DCF (see Section~\ref{sec:metrics}) and EER metrics.  Also illustrated are max min t-DCF results and corresponding attack identifier (AID) for each system (see Section~\ref{sec:analysis}).
    }
    \subfigure[Single systems]{\begin{tabular}{lccc}
         \toprule
         TID & min t-DCF & EER [\%] & \makecell{Max min t-DCF \\ (AID)} \\
         \midrule
        T45  & 0.1562	& 5.06 & 0.9905 (A17) \\
        T24  & 0.1655	& 4.04 & 0.8499 (A17) \\
        T39  & 0.1894	& 7.01 & 1.000 (A17) \\
        T01  & 0.1937	& 5.97 & 0.7667 (A17) \\
        T04  & 0.1939	& 5.74 & 0.7837 (A17) \\
         \midrule
         B01 & 0.2839 & 9.57 & 0.9901 (A17) \\
         B02 & 0.2605 & 8.09 & 0.6571 (A17) \\
         Perfect & 0.0627 & 0.0 & 0.4218 (A17) \\
         \bottomrule
    \end{tabular}
    \label{tab:la:single}
    }
    \subfigure[Primary systems]{
    \begin{tabular}{lccc}
         \toprule
         TID & min t-DCF & EER [\%] & \makecell{Max min t-DCF \\ (AID)} \\
         \midrule
         T05   &  0.0692  &	0.22   & 0.4418 (A17) \\
         T45   &   0.1104  & 1.86   & 0.7778 (A17) \\
         T60   &   0.1331  &	2.64   & 0.8803 (A17) \\
         T24   &   0.1518  & 3.45   & 0.8546 (A17) \\
         T50   &  0.1671  &	3.56   & 0.8471 (A17) \\
         \bottomrule
    \end{tabular}
    \label{tab:la:primary}
    }
    
    \label{tab:la}
\end{table}

\subsection{Primary systems}
\label{sec:la:primary}

The architectures of the top-5 primary systems are also illustrated in Fig.~\ref{fig:LA:illustration}. 
A short description of each follows: 

\noindent\textbf{T05:} A fusion of seven sub-systems, six of which derive spectral representations using the DFT, 
while the seventh uses the discrete cosine transform (DCT). Features are extracted using frame lengths of 256, 160, or 128 samples, frame overlaps of 100, 60, or 50 samples, and 256 or 160 point DFTs. 
The input features are sliced in 2D matrices with 256, 160, or 128 columns (frames).
All are based upon different neural network architectures: four upon MobileNetV2~\cite{sandler2018mobilenetv2}; two upon ResNet-50~\cite{he2016deep}; one upon DenseNet-121~\cite{huang2017densely}. 

\noindent\textbf{T45~\cite{lavrentyeva2019stc}:} A fusion of five sub-systems, 
including the LFCC-GMM baseline system (B1) and T45 single system. The remaining three sub-systems use LCNNs, each of which uses different features: LFCCs with CMVN; a log power spectrogram derived from the CQT; log power spectrogram derived from the DFT. All LCNN-based sub-systems use features extracted from the first 600 frames of each file. Sub-system scores are normalized according to the standard deviation of bona fide scores from the same sub-system before being fused using equal weights.

\noindent\textbf{T60~\cite{chettri2019ensemble}:} A fusion of seven sub-systems. 
Two sub-systems are based upon 128-component GMMs trained with either MFCC or \emph{inverted} MFCC (IMFCC)~\cite{chakroborty2007improved} features appended with delta and double-delta coefficients. The third sub-system is based upon the concatenation of 100-dimension i-vectors extracted from MFCC, IMFCC, CQCC~\cite{TODISCO2017516} features and sub-band centroid magnitude coefficient (SCMC)~\cite{kua2010investigation} features, and a support vector machine (SVM) classifier. 
The fourth sub-system is 
a CNN classifier operating on mean-variance normalized log DFT grams.
The remaining three sub-systems are based upon either mean-variance normalized Mel-scaled spectrograms or raw audio and either convolutional recurrent neural network (CRNN), Wave-U-Net~\cite{Stoller2018waveunet} or raw audio CNN classifiers. 
The NN-based sub-systems process a fixed number of feature frames or audio samples for each trial. 
Data for two attack conditions were excluded for training and used instead for validation and to stop learning.
Scores are combined according to logistic regression fusion.

\noindent\textbf{T24:} A fusion of two sub-systems: the T24 single system; a second sub-system using the same ResNet classifier but with CQCC-based features.
Scores are derived using single-layer neural networks before fusion (details unspecified). 

\noindent\textbf{T50~\cite{Yang2019}:} A fusion of six sub-systems all based on log-CQT gram features. Features are concatenated with a phase gram or compressed log-CQT gram obtained from a \emph{variational autoencoder} (VAE) trained on bona fide recordings only.  Three of the six classifiers use ResNet-18~\cite{he2016deep} classifiers, for one of which a standard i-vector is concatenated with the embedding layer of the network to improve generalizability. Two other classifiers use CGCNNs~\cite{dauphin2017language}. The last classifier (CGCRNN) incorporates bidirectional gated recurrent units~\cite{cho2014learning}.
Scores are combined by equal weight averaging.

\subsection{Results}
\label{sec:la:results}

A summary of results for the top-5 single and primary submissions is presented in Tables~\ref{tab:la:single} and~\ref{tab:la:primary} respectively.  Results for the two baseline systems appear in the penultimate two rows of Table~\ref{tab:la:single} whereas the last row shows performance for a perfect CM, \emph{i.e.}\ the ASV floor. 

In terms of the t-DCF, all of the top-5 single systems outperform both baselines by a substantial margin, with the best T45 system outperforming the B02 baseline by 40\% relative.  Both systems use LFCC features, whereas T45 uses a LCNN instead of a GMM-UBM classifier.  Even so, T01 and T04 systems, both also based upon standard cepstral features and GMM-UBM classifiers, are only slightly behind the better performing, though more complex systems. 

Four of the top-5 primary systems perform even better, with the best T05 primary system outperforming the best T45 single system by 56\% relative (73\% relative to B02).  The lowest min t-DCF of 0.0692 (T05 primary) is also only marginally above the ASV floor of 0.0627, showing that the best performing CM gives an expected detection cost that is close to that of a perfect CM.  The top-3 primary systems all combine at least 5 sub-systems.  All use diverse features, including both cepstral and spectral representations, with at least one DNN-type classifier.  Of note also are differences in performance between the same teams' primary and single systems.  Whereas the T05 primary system is first placed, the corresponding single system does not feature among the top-5 single systems, implying a substantial improvement through system combination. 
The first-placed T45 single system, however, is the second-placed primary system and, here, the improvement from combining systems is more modest.  The same is observed for T24 primary and single systems.

\section{Physical access scenario}
\label{sec:pa}

This section describes submissions to ASVspoof 2019 for the PA scenario and results.  It is organised in the same way as for the logical access scenario in Section~\ref{sec:la}.

\subsection{Single systems}
\label{sec:pa:single}

\begin{figure}[!t]
	\centering
	\includegraphics[trim=70 140 70 50,clip,width=1.0\linewidth]{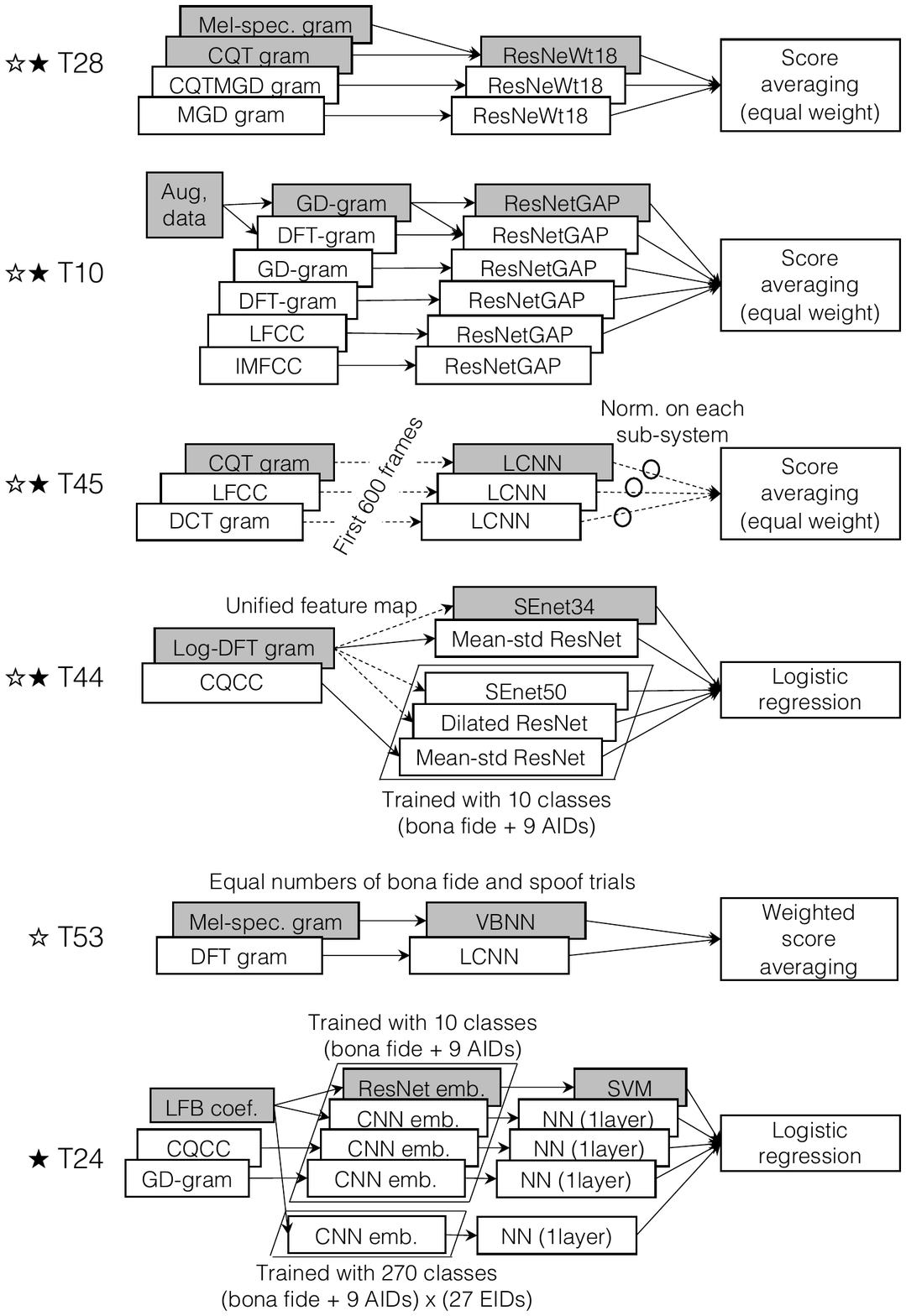}
	\caption{
    Illustration of top single (grey blocks) and primary system submissions for the PA scenario. $\largewhitestar$ and $\largeblackstar$ denote top-5 single and top-5 primary systems, respectively. 
	} 
	\label{fig:PA:illustration}
\end{figure}

The architectures of the top-5 single systems are illustrated in Fig.~\ref{fig:PA:illustration}  (grey blocks) in which systems are again labelled with the corresponding TID.
A short description of each follows:

\noindent\textbf{T28~\cite{cheng2019replay}:} Spectral features based upon the concatenation of Mel-grams and CQT-grams and 
a ResNetWt18 classifier based upon a modified ResNet18 architecture~\cite{He-ResNet-CVPR2016} where the second 3x3 convolution layer is split into 32 groups.  A $50\%$ dropout layer is used after pooling and the fully connected layer has a binary output.

\noindent\textbf{T10~\cite{Cai2019}:} Group delay (GD) grams~\cite{Francis-E2EaudioReplay-Iterspeech2018} with cepstral mean and variance normalisation (CMVN) with data augmentation via speed pertubation.  The classifier, referred to as ResNetGAP, is based upon a ResNet34 architecture~\cite{He-ResNet-CVPR2016} with 
a global average pooling (GAP) layer that transforms local features into 128-dimensional utterance-level representations which are then fed to a fully connected layer with softmax based cross-entropy loss.

\noindent\textbf{T45~\cite{lavrentyeva2019stc}:} Log power CQT-grams with a LCNN classifier 
which uses Kaiming initialization, additional batch normalizations, and angular softmax loss. 
In identical fashion to the T45 LA system, the PA system operates only upon the first 600 frames from each utterance and fuses scores by averaging.

\noindent\textbf{T44~\cite{Lai2019}:} Log-DFT grams with a unified feature map
and squeeze and excitation network (SEnet34)~\cite{Hu-SEnets-CVPR-2018} with a  ResNet34 backbone, in which each block aggregates channel-wise statistics (\emph{squeeze} operation) to capture channel-wise dependencies for an adaptive feature recalibration (\emph{excitation} operation) and binary training objective.

\noindent\textbf{T53:} Fixed-length log Mel grams (2048 frequency bins)
extracted from concatenated or truncated utterances 
and a variational Bayesian neural network (VBNN) using flipout~\cite{Wen-Flipout-arXiv2018} to decorrelate gradients within mini-batches.
Bona fide data is oversampled to make the bona fide and spoofed data balanced~\cite{Bialobrzeski2019}.

\begin{table}[!t]
    \centering
    \caption{As for Table.~\ref{tab:la} except for the PA scenario.  
    In contrast to the LA scenario, the worst case PA scenario is denoted by both the attack identifier (AID) and the environment identifier (EID). 
    Also illustrated here are min-tDCF results for hidden, real replay data (see Section~\ref{subsec:worstCase}) --- min t-DCF results for real replay data are computed using $C_0$, $C_1$, and $C_2$ terms derived for simulated replay data.  
    }
    \label{tab:PA-systems}
    \setlength{\tabcolsep}{4pt}
    \subfigure[Single systems\label{tab:PA-systems-single}]{
    \begin{tabular}{l*3c|*2c}
         \toprule
         & \multicolumn{3}{c}{Performance} & \multicolumn{2}{|c}{Hidden track} \\
         TID & \makecell{min \\ t-DCF} & \makecell{EER \\ (\%)} & \makecell{Max min t-DCF \\ (AID/EID)} & \makecell{min \\ t-DCF} & \makecell{EER \\ (\%)} \\
         \midrule
         T28 & 0.1470 & 0.52 & 0.2838 ($AA/acc$) & 0.5039 & 19.68 \\
         T10 & 0.1598 & 1.08 & 0.3768 ($AA/caa$) & 0.8826 & 37.04 \\
         T45 & 0.1610 & 1.23 & 0.2809 ($AA/acc$) & 0.7139 & 25.03 \\
         T44 & 0.1666 & 1.29 & 0.2781 ($AA,AC/acc$) & 0.7134 & 41.11 \\
         T53 & 0.1729 & 1.66 & 0.2852 ($BA/acc$) & 0.6379 & 32.64 \\
         \midrule
         B01 & 0.3476 & 11.04 & 1.0 ($BA/baa,caa,cac$) & 0.3855 & 12.73 \\
         B02 & 0.3481 & 13.54 & 1.0 ($BA/caa$) & 0.6681 & 29.44 \\
         Perfect & 0.1354 & 0.0 & 0.2781 ($AA/acc$) & - & - \\
         \bottomrule
    \end{tabular}
    \label{tab:pa:single}
    }
    \subfigure[Primary systems\label{tab:PA-systems-primary}]{
    \begin{tabular}{l*3c|*2c}
         \toprule
         & \multicolumn{3}{c}{Performance} & \multicolumn{2}{|c}{Hidden track} \\
         TID & \makecell{min \\ t-DCF} & \makecell{EER \\ (\%)} & \makecell{Max min t-DCF \\ (AID/EID)} & \makecell{min \\ t-DCF} & \makecell{EER \\ (\%)} \\
         \midrule
         T28 & 0.1437 & 0.39  & 0.2781 ($AA$,$AC/acc$) &  0.7160 & 30.74 \\
         T45 & 0.1460 & 0.54  & 0.2803 ($AA/acc$) &  0.6136 & 20.02 \\
         T44 &  0.1494 & 0.59  & 0.2781 ($AA$,$AC/acc$) &  0.6798 & 33.66 \\
         T10 &  0.1500 & 0.66  & 0.2781 ($AA$,$AC/acc$) &  0.7987 & 32.04  \\
         T24 &  0.1540 & 0.77  & 0.2781 ($AA$,$AC/acc$) &  0.9236 & 31.67 \\
         \bottomrule
    \end{tabular}
    \label{tab:pa:primary}
    }
    \label{tab:pa}
\end{table}

\subsection{Primary systems}
\label{sec:pa:primary}

The architectures of the top-5 primary systems are also illustrated in Fig.~\ref{fig:PA:illustration}.  
A short description of each follows: 
 
\noindent\textbf{T28~\cite{cheng2019replay}:} A fusion of three sub-systems, all ResNet variants referred to as ResNeWt18.
The first sub-system is the T28 single system and operates upon concatenated Mel and CQT grams.
The second operates upon a CQT modified group delay (CQTMGD) gram whereas the third operates directly upon the MGD gram (no CQT).  Scores are combined by equal weight averaging.

\noindent\textbf{T45~\cite{lavrentyeva2019stc}:} A fusion of 
three sub-systems with different frontends and a common  
LCNN backend.
The first sub-system is the T45 single system operating on CQT grams, while the other two use either LFCC or DCT grams.

\noindent\textbf{T44~\cite{Lai2019}:} A fusion of 
five sub-systems with either log-DFT gram or CQCC frontends and either squeeze and excitation network (SEnet) or ResNet based backends.
One sub-system is the T44 single system. 
Two are mean and standard deviation ResNets (Mean-std ReNets) for which the input feature sequences are transformed into a single feature vector through statistics pooling. 
Other classifiers receive fixed-size 2D feature matrices, referred to as unified feature maps~\cite{lai2019attentive}. 
All are either binary classifiers (\emph{i.e.}\ bona fide vs.\ spoof) or multi-class classifiers trained to predict the type of spoofing attack.
Scores are combined via logistic regression fusion.

\noindent\textbf{T10~\cite{Cai2019}:} A fusion of six sub-systems, all ResNet-based architectures with global average pooling (GAP) for utterance level aggregation. Two sub-systems, including the T10 single system, use data augmentation in the form of \emph{speed perturbation}~\cite{ko2015audio} applied to the raw signal.
Front-ends include 
group-delay (GD) gram, 
DFT gram, LFCCs and IMFCCs.  
Networks are configured as binary classifiers and trained with cross-entropy loss.  Scores coming from the bona fide unit for each sub-system are fused using equal weight score averaging.

\noindent\textbf{T24:} A fusion of five sub-systems using either LFB coefficients, CQCCs or GD-gram frontends and either
CNN or 
ResNet backends. 
Embeddings produced by the
ResNet system 
are length-normalised and classified using a weighted, two-class SVM.
Three of the CNN systems and the ResNet system are configured with 10 classes (combination of 9 AIDs and the bona fide class) whereas the other CNN system has 270 output classes (full combination of all EIDs, AIDs and bona fide class).  All use statistics pooling to obtain utterance-level representations from frame-level representations. Utterance-level embeddings are computed from the second-to-last fully connected layer in a similar manner to x-vector extraction. 
Except for the first sub-system, 
embeddings are 
processed with 
a single-layer neural network.
Scores are combined with logistic regression.

\subsection{Results}
\label{sec:pa:results}

A summary of results for the top-5 single and primary submissions is presented in
Tables~\ref{tab:pa:single} and~\ref{tab:pa:primary} respectively, with those for the two baseline systems and the ASV floor appearing in the last three rows of Table~\ref{tab:pa:single}.  

Just as is the case for the LA scenario, for the PA scenario all of the top-5 single systems outperform both baselines, again by a substantial margin.  In terms of the t-DCF, the best T28 system outperforms baseline B01 by 58\% relative.  In contrast to the LA scenario, however, all of the top-5 systems use spectral features rather than cepstral features and all also use DNN-type classifiers.  Of note also is the small gap in performance between the top-5 systems, and the use of data augmentation by only one of the top-5 systems, but not the top system.  The latter is, however, the only single system that uses concatenated Mel-gram and CQT-gram features.

In contrast to single systems, primary systems utilise both spectral and cepstral features, but again with exclusively DNN-type classifiers.  It seems, though, that system combination is less beneficial than for the LA scenario; primary system results for the PA scenario are not substantially better than those for single systems.  Perhaps unsurprisingly, then, teams with the best single systems are generally those with the best primary systems.  The best T28 primary system outperforms the best single system, also from T28, by only 2\% relative (59\% relative to B01).  Lastly, the lowest min t-DCF of 0.1437 (T28 primary) is only marginally above the ASV floor of 0.1354 showing, once again, that the best performing CM gives an expected detection cost that is close to that of a perfect CM.

\section{Analysis}
\label{sec:analysis}

This section aims to provide more in-depth analysis of results presented in Sections~3 and~4.  We 
report an analysis of generalisation performance which shows that results can be dominated by detection performance for
some so-called
worst-case spoofing attacks. 
Further analysis shows potential to improve upon fusion strategies 
through the use of more complementary sub-systems.

\begin{figure*}[!t]
    \centering
    \subfigure[LA]{
        \begin{tikzpicture}[font=\scriptsize]
            \begin{axis}[
                boxplot/draw direction=y,
                x axis line style={opacity=0},
                axis x line*=bottom,
                axis y line*=left,
                xticklabel style={align=center},
                ymajorgrids,
                ymin=0, ymax=1,
                xmin=0.35, xmax=6.65,
                xtick={1,2,3,4,5,6},
                xticklabels={
                    known\vphantom{/}, varied\vphantom{/}, unknown\vphantom{/},, varied w/o A17\hspace{2em}, A17\vphantom{/}
                },
                width=0.475\textwidth,
                height=3.5cm,
                ylabel={min t-DCF},
                xlabel={Acoustic model/waveform generator},
                cycle list={
                    black\\
                    black\\
                    black\\
                    white\\
                    black\\
                    black\\
                    },
                x tick style={opacity=0}
                ]
                \addplot+[boxplot, thick] table [row sep=\\, y index=0] {
                    data \\
                    0.063167493220615 \\
                    0.061610460189824 \\
                    0.062895716193285 \\
                    0.11615809750025 \\
                    0.130118730334672 \\
                };
                \addplot+[boxplot, thick] table [row sep=\\, y index=0] {
                    data \\
                    0.081580070601351 \\
                    0.148710843448586 \\
                    0.177304731251922 \\
                    0.19852205337109 \\
                    0.222153957661856 \\
                };
                \addplot+[boxplot, thick] table [row sep=\\, y index=0] {
                    data \\
                    0.056122191834694 \\
                    0.057190786480725 \\
                    0.07578876877045 \\
                    0.068341295467878 \\
                    0.07542421811391 \\
                };
                \addplot+[boxplot, thick] table [row sep=\\, y index=0] {
                    data \\
                    0.75 \\
                };
                \addplot+[boxplot, thick] table [row sep=\\, y index=0] {
                    data \\
                    0.072163031994864 \\
                    0.078681232316859 \\
                    0.112146257372834 \\
                    0.128101080829283 \\
                    0.187395611021668 \\
                };
                \addplot+[boxplot, thick] table [row sep=\\, y index=0] {
                    data \\
                    0.441842341269325 \\
                    0.777766725671074 \\
                    0.880335925462527 \\
                    0.854590149658446 \\
                    0.847140301830531 \\
                };
                 \addplot[draw=green!80!black,no markers,ultra thick] coordinates {(0.75,0.06161046018982388)(1.25,0.06161046018982388)};
                \addplot[draw=green!80!black,no markers,ultra thick] coordinates {(1.75,0.07434067076850695)(2.25,0.07434067076850695)};
                \addplot[draw=green!80!black,no markers,ultra thick] coordinates {(5.75,0.42176105731798846)(6.25,0.42176105731798846)};
                \addplot[draw=green!80!black,no markers,ultra thick] coordinates {(4.75,0.06536652773253905)(5.25,0.06536652773253905)};
                \addplot[draw=green!80!black,no markers,ultra thick] coordinates {(2.75,0.04998761893185581)(3.25,0.04998761893185581)};
            \end{axis}
        \end{tikzpicture}\label{fig:generalisability:la}}
        \hfill
    \subfigure[PA]{
        \begin{tikzpicture}[font=\scriptsize]
            \begin{axis}[
                boxplot/draw direction=y,
                x tick style={opacity=0},
                x axis line style={opacity=0},
                axis x line*=bottom,
                axis y line*=left,
                xticklabel style={align=center},
                ymajorgrids,
                ylabel={min t-DCF},
                ymin=0, ymax=.3,
                xtick={1,2,3,4,5,6},
                xmin=0.35, xmax=6.65,
                xticklabels={
                    $a$: low\vphantom{/}, 
                    $b$: medium\vphantom{/},
                    $c$: high\vphantom{/},,
                    $c$: high w/o $acc$\hspace{2em},
                    $acc$\vphantom{/}
                },
                width=0.475\textwidth,
                height=3.5cm,
                xlabel={$T60$ reverberation},
                cycle list={
                    black\\
                    black\\
                    black\\
                    white\\
                    black\\
                    black\\
                    },
                ]
                \addplot+[boxplot, thick] table [row sep=\\, y index=0] {
                    data \\
                    0.080628200220292 \\
                    0.086650934814607 \\
                    0.101228074451136 \\
                    0.100302066204483 \\
                    0.099463095093004 \\
                };
                \addplot+[boxplot, thick] table [row sep=\\, y index=0] {
                    data \\
                    0.130972365627134 \\
                    0.133783862915764 \\
                    0.132051631533082 \\
                    0.132968304064065 \\
                    0.133935101438544 \\
                };
                \addplot+[boxplot, thick] table [row sep=\\, y index=0] {
                    data \\
                    0.213713355328971 \\
                    0.213646210321407 \\
                    0.210886695826389 \\
                    0.212892474587804 \\
                    0.21282094571662 \\
                };
                \addplot+[boxplot, thick] table [row sep=\\, y index=0] {
                    data \\
                    0.75 \\
                };
                \addplot+[boxplot, thick] table [row sep=\\, y index=0] {
                    data \\
                    0.158834225647914 \\
                    0.160715890532437 \\
                    0.162605600913577 \\
                    0.160301175513345 \\
                    0.165919888485768 \\
                };
                \addplot+[boxplot, thick] table [row sep=\\, y index=0] {
                    data \\
                    0.274747865930719 \\
                    0.274915865035206 \\
                    0.274243868617259 \\
                    0.274243868617259 \\
                    0.274579866826232 \\
                };
                \addplot[draw=green!80!black,no markers,ultra thick] coordinates {(0.75,0.06668136427148051)(1.25,0.06668136427148051)};
                \addplot[draw=green!80!black,no markers,ultra thick] coordinates {(1.75,0.12544469930403196)(2.25,0.12544469930403196)};
                \addplot[draw=green!80!black,no markers,ultra thick] coordinates {(2.75,0.20985023394189883)(3.25,0.20985023394189883)};
                \addplot[draw=green!80!black,no markers,ultra thick] coordinates {(5.75,0.2742438686172589)(6.25,0.2742438686172589)};
                \addplot[draw=green!80!black,no markers,ultra thick] coordinates {(4.75,0.15658161286163952)(5.25,0.15658161286163952)};
            \end{axis}
        \end{tikzpicture}\label{fig:generalisability:pa}}
    \caption{Illustrations of generalisation performance for the top-5 primary systems for the LA scenario (a) and the PA scenario (b), estimated using evaluation set data. For the LA scenario, box plots illustrate performance decomposed across known, varied and unknown attacks. For the scenario, they illustrate performance decomposed across low, medium and high $T60$ reverberation categories.  For all plots, the green profiles signify corresponding ASV floors (performance for a perfect CM). 
    The two right-most box plots in each case indicate performance for varied attacks without the worst case AID (LA) and high $T60$ reverberation without the worst case AID/EID (PA) and then for the worst case scenarios on their own (see Section~\ref{subsec:worstCase}).
    }
    \label{fig:generalisability}
\end{figure*}
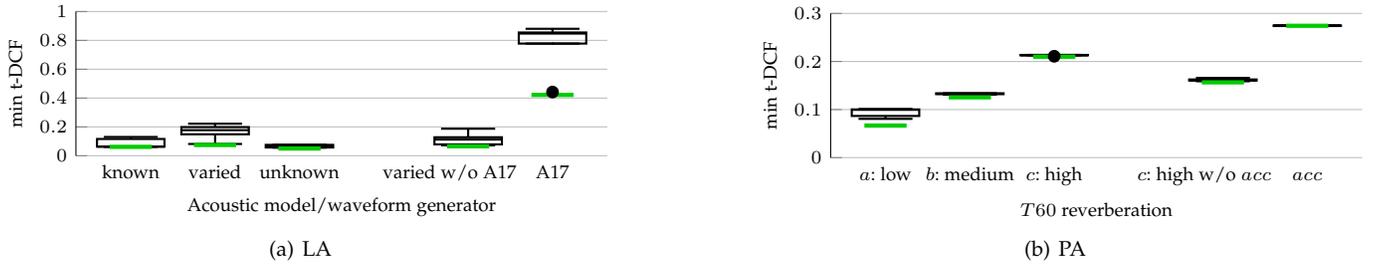

\subsection{Generalisation to unseen attacks}
\label{subsec:generalisation}

Since its inception, ASVspoof has prioritised strategies to promote the design of generalised CMs that perform reliably in the face of spoofing attacks not seen in training data.
For ASVspoof~2019, the LA evaluation set features TTS and VC spoofing attacks generated with algorithms for which some component (\emph{e.g.}\ the acoustic model or the waveform generator) is different to those used in generating spoofing attacks in the training and development sets.  The situation is different for the PA scenario.  While the full set of attack identifier~(AID) and environment identifier~(EID) \emph{categories} (see last two paragraphs of Section~\ref{sec:pa_description}) are seen in all three data sets, the \emph{specific} AIDs and EIDs in each are different (while the categories are the same, no specific attack or room configuration appears in more than one set).

A view of generalisation performance for the top-5 LA and PA primary system submissions is illustrated in Figures~\ref{fig:generalisability:la} and~\ref{fig:generalisability:pa} respectively.  
For the LA scenario, the three left-most box plots depict performance in terms of the min t-DCF for: known attacks (attacks that are identical to those seen in training and evaluation data); varied attacks (attacks for which either the acoustic model or waveform generator is identical to those of attacks in the training and development data); wholly unknown attacks (attacks for which both components are unseen).
Interestingly, while performance for unknown attacks is not dissimilar to, or even better than that for known attacks, there is substantial variability in performance for varied attacks.  This observation is somewhat surprising since, while systems appear to generalise well to unknown attacks, they can fail to detect others that are generated with only variations to known attack algorithms.
This can mean either that the unknown attack algorithms produce artefacts that are not dissimilar to those produced with known attacks, or that there is some peculiarity to 
the varied attacks.  The latter implies that knowledge of even some aspects of an attack is of little use in terms of CM design; CMs are over-fitting and there is potential for them to be overcome with perhaps even only slight adjustments to an attack algorithm.  Reassuringly, however, as already seen from results in Table~\ref{tab:la} and by the green profiles to the base of each box plot in Fig.~\ref{fig:generalisability:la} which illustrate the ASV floor, some systems produce min t-DCFs close to that of a perfect CM.

A similar decomposition of results for the PA scenario is illustrated in Fig.~\ref{fig:generalisability:pa}, where the three left-most box plots show performance for low, medium and high $T60$ reverberation categories, the component of the EID which was observed to have the greatest influence on performance.  In each case results are pooled across the other AID and EID components, namely the room size $S$ and the talker-to-ASV distance $D_\text{s}$.  Results show that, as the level of reverberation increases, the min t-DCF increases.  However, comparisons of each box plot to corresponding ASV floors show that
the degradation is not caused by the CM, the performance of which improves with increasing reverberation; replay attacks propagate twice in the same environment and hence reverberation serves as a cue for replay detection.  The degradation is instead caused by the performance of the ASV system; the gap between the min t-DCF and the ASV floor decreases with increasing $T60$ and, for the highest level of reverberation, the min t-DCF is close to the ASV floor.
This observation also shows that the effect of high reverberation dominates the influence of the room size and the talker-to-ASV distance.

From the above, it is evident that min t-DCF results are dominated by performance for some worst case attack algorithms (LA) or some worst case environmental influence (PA).  Since an adversary could exploit knowledge of such worst case conditions in order to improve their chances of manipulating an ASV system, it is of interest to examine not just the pooled min t-DCF, but also performance in such worst case scenarios.

\subsection{Worst case scenario}
\label{subsec:worstCase}

The worst case or maximum of the minimum (max min) t-DCFs (see Section~\ref{sec:metrics}) for the top-5 single and primary systems in addition to the baseline systems are shown in Tables~\ref{tab:la} and~\ref{tab:pa}. 
For the LA scenario, the worst case attack identifier (AID) is A17, a VC system that combines a VAE acoustic model with direct waveform modification~\cite{kobayashi2018intra}. 
While
the best individual result for A17 is obtained by the best performing primary system
the max min t-DCF is over 6 times higher than the min t-DCF.
The lowest max min t-DCF for single systems is 
that of baseline B02.  While this result (0.6571) is not substantially worse than the lowest max min t-DCF for primary systems (0.4418), it suggests that the fusion of different CMs may help to reduce the threat in the worst case scenario.  The two, right-most box plots in Fig.~\ref{fig:generalisability:la} show performance for varied attacks without attack A17, and then for attack A17 on its own, both for the top-5 performing primary LA systems.  A17 is a varied attack and is the single attack that accounts in large part for the differences between the box plots for varied and known/unknown attacks described in Section~\ref{subsec:generalisation}.

Performance for the PA scenario is 
influenced by both the attack (AID) and the environment (EID).  Excluding baselines, all but two systems struggle most for $acc$ EIDs with
small rooms~($a$), high $T60$ reverberation times~($c$) and large talker-to-ASV distances $D_s$~($c$),
and either the $AA$ or $AC$ AID where
recordings are captured 
in close proximity to the talker.
The two, right-most box plots in Fig.~\ref{fig:generalisability:pa} show performance for high $T60$ reverberation time without the $acc$ EID and then for the $acc$ EID on its own, both for the top-5 performing primary PA systems.  
Worst case max min t-DCFs 
are substantially higher than pooled min t-DCFs. Even so, it appears that the greatest influence upon tandem performance in the case of PA is within the system designer's control; the environment in which CM and ASV systems are installed should have low reverberation. 
Individual system results shown in Table~\ref{tab:pa} show 
that 
a single, rather than a primary system gives almost the best pooled min t-DCF (0.1470 cf.\ 0.1437).  This observation suggests that fusion techniques were not especially successful for the PA scenario.  We expand on this finding next.

\begin{figure}[!t]
	\centering
	\includegraphics[width=.9\linewidth]{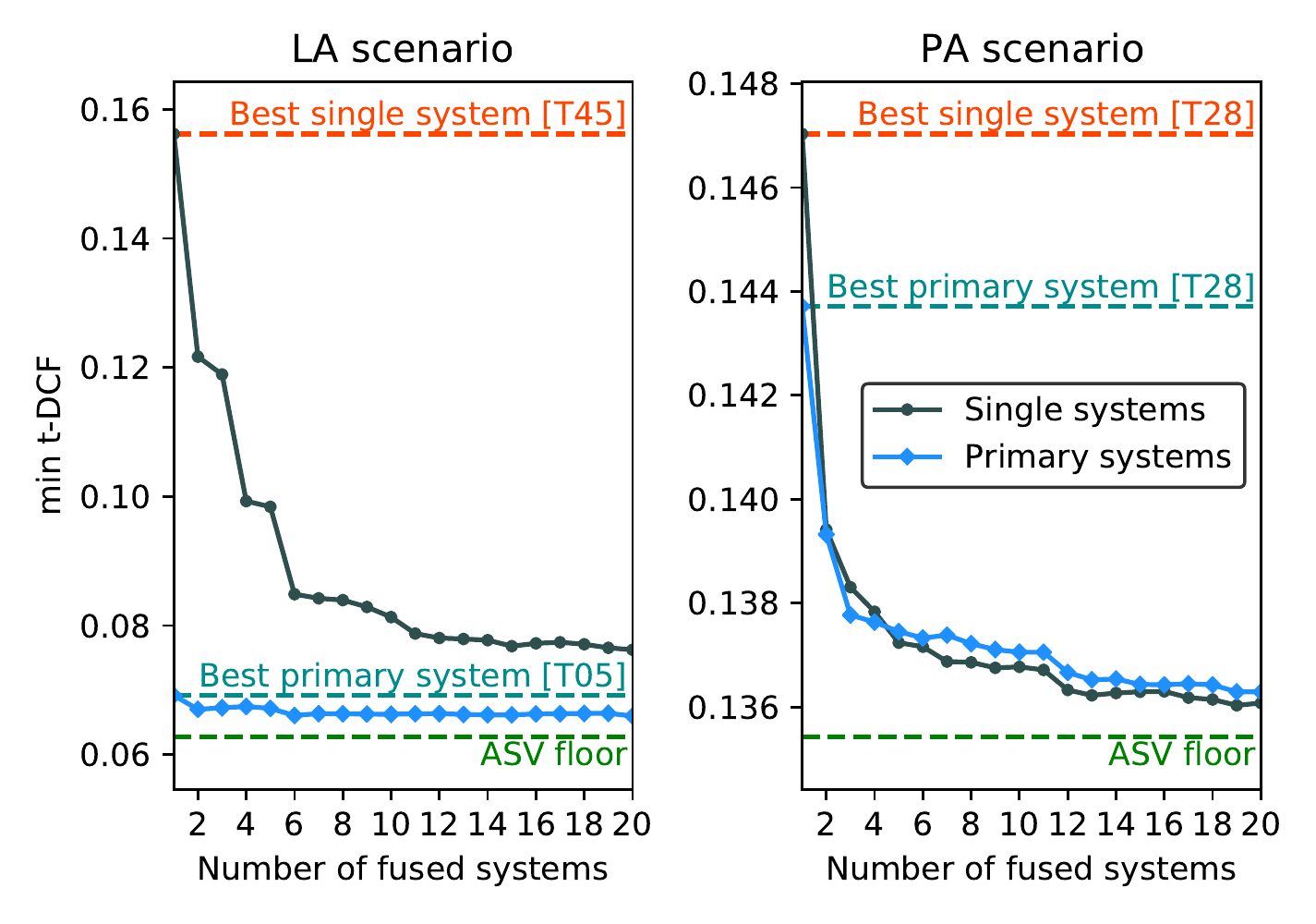}
	\caption{min t-DCF results for oracle fusion performed with evaluation set scores for the top-20 performing systems for the LA scenario (left) and PA scenario (right). System T08, which returned problematic score distributions, was excluded in compution of results for the LA scenario.}
	\label{fig:oracle_fusion}
\end{figure}

\subsection{Fusion performance}
\label{subsec:fusion}

From the treatment of results presented in Sections~\ref{sec:la:results} and~\ref{sec:pa:results}, we have seen already that fusion seems more beneficial for the LA scenario than for the PA scenario; the best performing single and pimary LA systems give min t-DCFs of 0.1562 and 0.0692 respectively, whereas the best performing single and primary PA systems give similar min t-DCFs of 0.1470 and 0.1437 respectively.

For the LA scenario, we note that the best performing T45 \emph{single} system still outperforms the fifth-placed T50 \emph{primary} system.  The architectures of the top-4 primary systems might then suggest that the benefit from fusion requires substantial investment in front-end feature engineering in addition to the careful selection and optimisation of the classifier ensemble. By way of example, the top-ranked T05 primary LA system uses DFT grams with different time-frequency resolutions, and three different classifiers in the shape of MobileNet, DenseNet, and a large ResNet-50.  In addition, two out of the seven sub-systems take into consideration the ratio of bona fide and spoofed samples observed in training.

Even if different front-end combinations are among the top-performing primary PA systems, we do not see the same diversity in the classifier ensembles.  We hence sought to investigate whether this lack of diversity could explain why fusion appears to have been less beneficial for the PA scenario. Using logistic regression~\cite{Brummer-deVilliers-BOSARIS-Binary-Scores-AGNITIO-Research-2011}, we conducted oracle fusion experiments for LA and PA evaluation datasets using the scores generated by the top-20 primary and single systems. In each case the number of systems in the fusion was varied between 2 and~20. 

Results are illustrated in Figure.~\ref{fig:oracle_fusion}.  Also illustrated in each plot is the min t-DCF for the best performing primary and single systems in addition to the ASV floor, \emph{i.e.}\ a prefect CM that makes no errors such that the only remaining errors are made by the ASV system. There are stark differences between the two challenge scenarios.  For the LA scenario, the best-performing T05 primary system (left-most, blue point) obtains nearly perfect results (performance equivalent to the ASV floor) and the fusion of multiple 
primary 
systems improves only marginally upon performance.  While the fusion of multiple single systems (black profile) leads to considerably better performance, even 
though
the fusion of 20 single systems fails to improve upon the best T05 primary system.  

As we have seen already, there is little difference between primary and single systems for the PA scenario. In addition, the performance of the best individual primary and single systems is far from that of the ASV floor; there is substantial room for improvement. Furthermore, the fusion of both primary and single systems gives substantially improved performance, to within striking distance of the ASV floor.  Fusion of only the best two single systems results in performance that is superior to the best T28 primary system. There was significant scope for participants to improve performance for the PA condition using the fusion of even only a small number of diverse sub-systems; it seems that those used by participants lack complimentarity.

\subsection{Progress}
\label{subsec:trends}

\label{sec:progress}

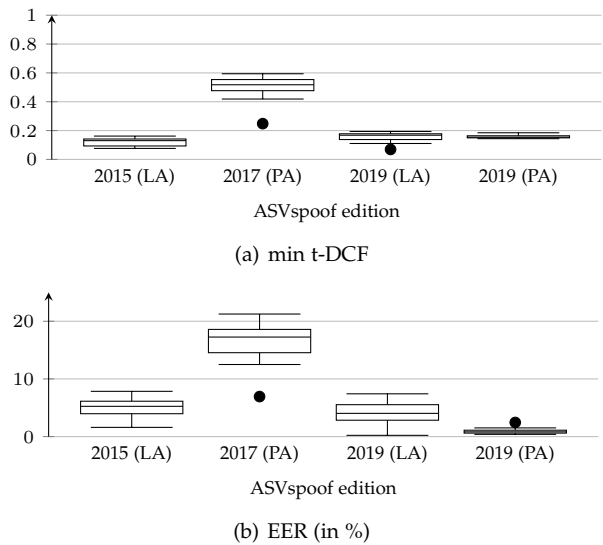
\begin{figure}[!t]
    \centering
    \subfigure[min t-DCF]{\begin{tikzpicture}[font=\scriptsize]
        \begin{axis}[
            boxplot/draw direction=y,
            x axis line style={opacity=0},
            axis x line*=bottom,
            axis y line=left,
            xticklabel style={align=center},
            ymajorgrids,
            ymin=0, ymax=1,
            xmin=0.35, xmax=4.65,
            xtick={1,2,3,4},
            xticklabels={
                {2015 (LA)},
                {2017 (PA)},
                {2019 (LA)},
                {2019 (PA)}
            },
            width=1\linewidth,
            height=3.5cm,
            ylabel={},
            xlabel={ASVspoof edition},
            cycle list={
                black\\
                },
            x tick style={opacity=0}
            ]
            \addplot+[boxplot] table [row sep=\\, y index=0] {
                data \\
                0.092 \\
                0.0762 \\
                0.1283 \\
                0.0896 \\
                0.1318 \\
                0.1462 \\
                0.1422 \\
                0.1618 \\
                0.0969 \\
                0.1392 \\
            };
            \addplot+[boxplot] table [row sep=\\, y index=0] {
                data \\
                0.248 \\
                0.4187 \\
                0.4736 \\
                0.4858 \\
                0.5345 \\
                0.5602 \\
                0.5263 \\
                0.5606 \\
                0.594 \\
                0.509 \\
            };
            \addplot+[boxplot] table [row sep=\\, y index=0] {
                data \\
                0.0692 \\
                0.1104 \\
                0.1331 \\
                0.1518 \\
                0.1671 \\
                0.1684 \\
                0.1752 \\
                0.1785 \\
                0.1875 \\
                0.1939 \\
            };
            \addplot+[boxplot] table [row sep=\\, y index=0] {
                data \\
                0.1437 \\
                0.146 \\
                0.1494 \\
                0.15 \\
                0.154 \\
                0.1543 \\
                0.1585 \\
                0.1657 \\
                0.1676 \\
                0.1847 \\
            };
        \end{axis}
    \end{tikzpicture}}
    \subfigure[EER (in \%)]{\begin{tikzpicture}[font=\scriptsize]
        \begin{axis}[
            boxplot/draw direction=y,
            x axis line style={opacity=0},
            axis x line*=bottom,
            axis y line=left,
            xticklabel style={align=center},
            ymajorgrids,
            ymin=0, ymax=25,
            xmin=0.35, xmax=4.65,
            xtick={1,2,3,4},
            xticklabels={
                {2015 (LA)},
                {2017 (PA)},
                {2019 (LA)},
                {2019 (PA)}
            },
            width=1\linewidth,
            height=3.5cm,
            ylabel={},
            xlabel={ASVspoof edition},
            cycle list={
                black\\
                },
            x tick style={opacity=0}
            ]
            \addplot+[boxplot] table [row sep=\\, y index=0] {
                data \\
                1.6 \\
                2.56 \\
                4.9 \\
                3.66 \\
                5.52 \\
                6.12 \\
                6.14 \\
                6.66 \\
                4.98 \\
                7.85 \\
            };
            \addplot+[boxplot] table [row sep=\\, y index=0] {
                data \\
                6.94 \\
                12.5 \\
                14.39 \\
                15.02 \\
                16.56 \\
                17.96 \\
                18.72 \\
                18.19 \\
                21.25 \\
                20.32 \\
            };
            \addplot+[boxplot] table [row sep=\\, y index=0] {
                data \\
                0.22 \\
                1.86 \\
                2.64 \\
                3.45 \\
                3.56 \\
                4.5 \\
                7.42 \\
                4.92 \\
                6.14 \\
                5.74 \\
            };
            \addplot+[boxplot] table [row sep=\\, y index=0] {
                data \\
                0.39 \\
                0.54 \\
                0.59 \\
                0.66 \\
                0.77 \\
                0.88 \\
                0.96 \\
                1.16 \\
                1.51 \\
                2.45 \\
            };
        \end{axis}
    \end{tikzpicture}}
    \caption{An illustration of performance for top-10 submission to the three ASVspoof challenge editions: 2015, 2017 and 2019.  Results are shown in terms of both min-DCF and EER.
    }
    \label{fig:comparison:progress}
\end{figure}

Fig.~\ref{fig:comparison:progress} shows box plots of performance for the top-10 systems for the three challenge editions: ASVspoof~2015 (LA), ASVspoof~2017 (PA), and ASVspoof~2019 (LA+PA). 
Comparisons should be made with caution; each database has different partitioning schemes or protocols and was created with different spoofing attacks.  Furthermore, while the ASVspoof~2017 database was created from the re-recording of a source database, the ASVspoof~2019 PA database was created using simulation and, while systems developed for 2015 and 2017 editions were optimised for the EER metric, those developed for 2019 may have been optimised for the new t-DCF metric.  Accordingly, Fig.~\ref{fig:comparison:progress} shows results in terms of both EER and min t-DCF.

Results for 2015 and 2019 LA databases shows that progress in anti-spoofing 
has
kept apace with progress in TTS and VC research, including neural network-based waveform modelling techniques including WaveNet~\cite{oord2016wavenet}; 
EERs and min t-DCFs are similar,
despite the use of state-of-the-art neural acoustic and waveform models to generate spoofing attacks in the ASVspoof 2019 database.
Results
for 2017 and 2019 PA databases 
seemingly 
show significant progress, with both EERs and min t-DCFs dropping by substantial margins, though
improvements are 
likely caused by database differences.
The 
2017 database contains both additive background noise and convolutional channel noise, artefacts stemming from the source database rather than being caused by replay spoofing, whereas the 2019 database contains neither.
EER results for the ASVspoof~2019 database are substantially lower than those for any of the other three databases,
indicating that
results reflect 
acoustic environment 
effects
upon the ASV system, rather than upon CM systems.  While this finding is encouraging
differences between 2017 and 2019 PA results show that additive noise might have a considerable impact on performance.  These issues are expanded upon next.

\section{Results for real replay recordings}
\label{sec:realReplay}

Results for simulated replay data were compared to results for 
real replay data\footnote{\url{https://www.asvspoof.org/database}} that were
concealed within the PA database.
This data, results for which were excluded from challenge scoring and ranking, is not described in~\cite{wang2020asvspoof}. 
Accordingly, a brief description is provided here.  Real replay data was recorded 
in $3$ different rooms
with two different talker-to-ASV distance categories $D_s$ 
and
in conditions 
equivalent to two different EID categories: a small meeting room (equivalent EIDs of $aaa$ and $aac$); a large office (equivalent EIDs of $bba$ and $bbc$); a small/medium office (equivalent to EIDs of $cca$ and $ccc$). 
Recordings were captured 
using high or low quality capture devices, whereas replay data were recorded using various acquisition devices before presentation to the ASV microphone using various presentation devices.  Both recording and presentation devices were of quality equivalent to either $B$ or $C$ categories. 
Data were collected from 26 speakers, each of whom provided 15 utterances selected at random from the same set of phonetically-balanced TIMIT phrases as the VCTK source data.
This setup gave 540 bona fide utterances and 2160 replay utterances.

In contrast to 
simulated data,
real replay data
contains additive, ambient noise.  
Differences between simulation and the collection of real replay data also imply that
consistent trends between 
results for the two datasets
cannot be expected.  
The objective of this analysis is to expose consistencies or discrepancies in results derived
between simulated and real data in terms of the t-DCF,
or to determine whether the use of simulated data leads to the design of CMs that perform well when tested with real data.
The two right-most columns of Table~\ref{tab:PA-systems} show min t-DCF and EER results for the baselines and top-5 single and primary systems.  
In general, there are substantial differences.
Among the top-5 systems considered, the best t-DCF result for real data of 0.3855 is obtained by the B01 baseline. 
This observation
suggests that CMs are over-fitting to simulated data,
or that 
CMs 
lack robustness to background noise.
This possibility seems likely; we observed 
greater consistency in results for
simulated replay data and real replay data recorded in quieter rooms.  One other explanation lies in the 
relation between additive noise and the ASV floor.
Results for synthetic data are dominated by the ASV floor,
whereas those for real data are dominated 
by the impact of additive noise.
Whatever the reason for these differences, their scale is cause for concern.
Some plans to address this issue are outlined in our thoughts for future directions.

\section{Future directions}
\label{sec:future}

Each 
ASVspoof challenge
raises new research questions 
and exposes ways in which the challenge 
can be 
developed and
strengthened.  A selection of 
these is presented here.

\subsection*{Additive noise and channel variability}

It
is likely that 
ambient and channel noise will degrade CM performance, thus it will be imperative to 
study the impact of such nuisance variation in future editions, \emph{e.g.} as in~\cite{Gong2019}.
Even if such practice is generally frowned upon, the \emph{artificial} addition of 
nuisance variation in a controlled fashion
may be appropriate at this stage.  
LA scenarios generally involve some form of telephony, \emph{e.g.}\ VoIP. 
Coding and compression
effects are readily simulated to some extent.  In contrast, the consideration of \emph{additive} noise is potentially more complex for it 
influences speech \emph{production}, \emph{e.g.}\ the Lombard reflex~\cite{LombardReflex}. 
The simulation of additive noise is then generally undesirable.
An appropriate strategy to address these issues in future editions of ASVspoof demands careful reflection.

\subsection*{Quality of TTS/VC training data}

For ASVspoof~2019, all TTS and VC systems 
were trained with data recorded in 
benign acoustic conditions.
This setup is obviously not representative of \emph{in-the-wild} scenarios 
where an adversary could acquire only relatively noisy training or adaptation data.
Future editions of ASVspoof 
should hence
consider TTS and VC attacks generated with 
more realistic data. 
Such attacks may be less effective in fooling ASV system, and may also be more easily detectable.

\subsection*{Diversified spoofing attacks}

ASVspoof presents an arguably naive view of potential spoofing attacks.
Future editions should consider more diversified attacks, \emph{e.g.}\ impersonation~\cite{Vestman2020-mimicry}, attacks by twins or siblings~\cite{kamble2020advances}, non-speech~\cite{Alegre2012-artificial} or adversarial attacks~\cite{Kreuk2018-adversarial,Li2020-adversarial-ivector,Rohan2020-attacker-perspective} and attacks that are injected into specific regions of the speech signal rather than the entire utterance.  One can also imagine blended attacks whereby, for instance, replay attacks are launched in an LA scenario, or replay attacks in a PA scenario are performed with speech data generated using TTS or VC systems. 

\subsection*{Joint CM+ASV score calibration}

With
the 2019 edition transitioned to an ASV-centric form of assessment with the min t-DCF metric
there are now not two, but three decision outcomes: target, non-target (both bona fide) and spoof.  The existing approaches to calibrate binary classification scores are then no longer suitable. Future work could hence investigate approaches to joint CM+ASV system optimisation and calibration.

\subsection*{Reproducibility}

Anecdotal evidence shows that some ASVspoof results are un-reproducible.  While it is not our intention to enforce reproducibility -- doing so may deter participation -- it is nonetheless something that we wish to promote.  One strategy is to adopt the reviewing of system descriptions, either by the organisers or by fellow ASVspoof participants, or the reporting of system descriptions according to a harmonised format.  
Such a harmonised reporting format should include details of the fusion scheme and weights.  This policy, together with a requirement for the submission of scores for each system in an ensemble, would also allow a more fine-grained study of fusion strategies and system complementarity.

\subsection*{Explainability}

Explainability is a topic of growing importance in almost any machine learning task and is certainly lacking sufficient attention in the field of anti-spoofing.  
While the results reported in this paper show promising potential to detect spoofing attacks, we have learned surprisingly little about the artefacts or the cues that distinguish bona fide from spoofed speech.  Future work which reveals these cues may be of use to the community in helping to design better CMs.

\subsection*{Standards}

The t-DCF 
metric adopted for ASVspoof 2019
does not meet security assessment standards, in particular the so-called \emph{Common Criteria for Information Technology Security Evaluation}~\cite{CommonCriteria-CC-SecurityAssuranceComponents-2017,ISO-IEC-19989-1-DIS-2019,ISO-IEC-19989-3-DIS-2019}.  Rather than quantifying a probabilistic meaning of some attack likeliness, common criteria are based upon a category-based points scheme in order to determine a so-called \emph{attack potential}. 
This reflects
\emph{e.g.}\ the equipment, expertise and time required to mount the attack and the knowledge required of the system under attack. 
The rigour in common criteria is that each category of attack potential then demands hierarchically higher \emph{assurance components} in order to meet \emph{protection profiles} that express \emph{security assurance requirements}. 
In the future, it may prove beneficial to explore the gap between the common criteria and the t-DCF.  To 
bridge this gap pragmatically, we need to determine the attack potentials for ASVspoof, asking ourselves: 1)~\emph{How long does it take to mount a given attack?}; 2)~\emph{What level of expertise is necessary?}; 3)~\emph{What resources (data or computation) are necessary to execute it?};
4)~\emph{What familiarity with the ASV system is needed?} Clearly, providing the answers to these questions is far from being straightforward.

\subsection*{Challenge model}

The organisation of ASVspoof has developed into a demanding, major organisational challenge involving the coordination of 6 different organising institutes and 19 different data contributors for the most recent edition.  While the organisers enjoy the support of various different national research funding agencies, it is likely that we will need to attract additional industrial, institutional or public 
funding to support the initiative 
in the future.  To this end, we are liaising with the Security and Privacy in Speech Communications~(SPSC), the Speaker and Language Characterisation~(SpLC) and Speech Synthesis~(SynSig)
Special Interest Groups~(SIGs) of the International Speech Communication Association~(ISCA) regarding a framework with which to support the initiative in the longer term.

With regards the format, the organising team committed to making ASVspoof 2019 the last edition to be run as a special session at INTERSPEECH.  With anti-spoofing now featuring among the Editor’s Information Classification Scheme (EDICS) and topics of major conferences and leading journals/transactions, it is time for ASVspoof to make way for more genuinely `special' topics.  Accordingly, we will likely transition in the future to a satellite workshop format associated with an existing major event, such as INTERSPEECH. 

\section{Conclusions} 

ASVspoof 2019 is the third in the series of anti-spoofing challenges for automatic speaker verification.  It was the first to consider both logical and physical access scenarios in a single evaluation and the first to adopt the tandem detection cost function as the default metric and also brought a series of additional advances with respect to the 2015 and 2017 predecessors.  With the database and experimental protocols described elsewhere, the current paper describes the challenge results, findings and trends, with a focus on the top-performing systems for each scenario.

Results reported in the paper are encouraging and point to advances in countermeasure performance.  For the logical access scenario, reliability seems to have kept apace with the recent, impressive developments in speech synthesis and voice conversion technology, including the latest neural network-based waveform modelling techniques. 
For the physical access scenario, countermeasure performance is stable across diverse acoustic environments.  Like most related fields in recent years, the 2019 edition was marked with a shift towards deep architectures and ensemble systems that brought substantial improvements in performance, though more so for the logical access scenario than the physical access counterpart.  There seems to have been greater diversity among each teams' ensemble systems for the former, while there is evidence that those for the latter suffer from over-fitting.  In both cases, however, \emph{tandem} systems exhibit high detection costs under specific conditions: either specific attack algorithms or specific acoustic environments with costs stemming from either countermeasures or automatic speaker verification systems. 

Many challenges remain.  Particularly for the physical access scenario, though likely also for the logical access scenario, countermeasure performance may degrade in real-world conditions characterised by nuisance variation such as additive noise.  Results for real replay data with additive noise show substantial gaps between results for simulated, noise-free data.  It is furthermore reasonable to assume that performance will also be degraded by channel variability as well as any mismatch in the training data used to generate spoofing attacks.

Future editions of ASVspoof will also consider greater diversification in spoofing attacks and blended attacks whereby speech synthesis, voice conversion and replay attack strategies are combined.  Countermeasure optimisation strategies also demand further attention now that they are assessed in tandem with automatic speaker verification systems.  
Future editions 
demand greater efforts to promote reproducibility and explainability, as well as some reflection on the gap between ASVspoof and standards such as the common criteria.  Lastly, the paper outlines our plans to adopt a satellite event format, rather than a special session format for the next edition, tentatively, ASVspoof~2021.

\section*{Acknowledgements}
The ASVspoof 2019 organisers thank the following for their invaluable contribution to the LA data collection effort -- Lauri Juvela, Paavo Alku, Yu-Huai Peng, Hsin-Te Hwang, Yu Tsao, Hsin-Min Wang, Sebastien Le Maguer, Markus Becker, Fergus Henderson, Rob Clark, Yu Zhang, Quan Wang, Ye Jia, Kai Onuma, Koji Mushika, Takashi Kaneda, Yuan Jiang, Li-Juan Liu, Yi-Chiao Wu, Wen-Chin Huang, Tomoki Toda, Kou Tanaka, Hirokazu Kameoka, Ingmar Steiner, Driss Matrouf, Jean-Francois Bonastre, Avashna Govender, Srikanth Ronanki, Jing-Xuan Zhang and Zhen-Hua Ling.  We also extend our thanks to the many researchers and teams who submitted scores to the ASVspoof 2019 challenge.  Since participants are assured of anonymity, we regret that we cannot acknowledge them here by name.

\appendices

\ifCLASSOPTIONcaptionsoff
  \newpage
\fi

\bibliographystyle{IEEEtran}
\bibliography{IEEEabrv,main.bib}

\begin{IEEEbiography}[{\includegraphics[width=1in,height=1.25in,clip,keepaspectratio]{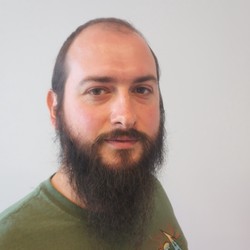}}]{Andreas Nautsch} is with the Audio Security and Privacy research group (EURECOM). He received the doctorate from Technische Universit\"{a}t Darmstadt in 2019. From 2014 to 2018, he was with the da/sec Biometrics and Internet-Security research group (Hochschule Darmstadt) within the German National Research Center for Applied Cybersecurity. He received B.Sc. and M.Sc. degrees from Hochschule Darmstadt (dual studies with atip GmbH) respectively in 2012 and 2014. He served as an expert delegate to ISO/IEC and as project editor of the ISO/IEC 19794-13:2018 standard. Andreas serves currently as associate editor of the EURASIP Journal on Audio, Speech, and Music Processing, and is a co-initiator and secretary of the ISCA Special Interest Group on Security \& Privacy in Speech Communication.
\end{IEEEbiography}
\begin{IEEEbiography}[{\includegraphics[width=1in,height=1.25in,clip,keepaspectratio]{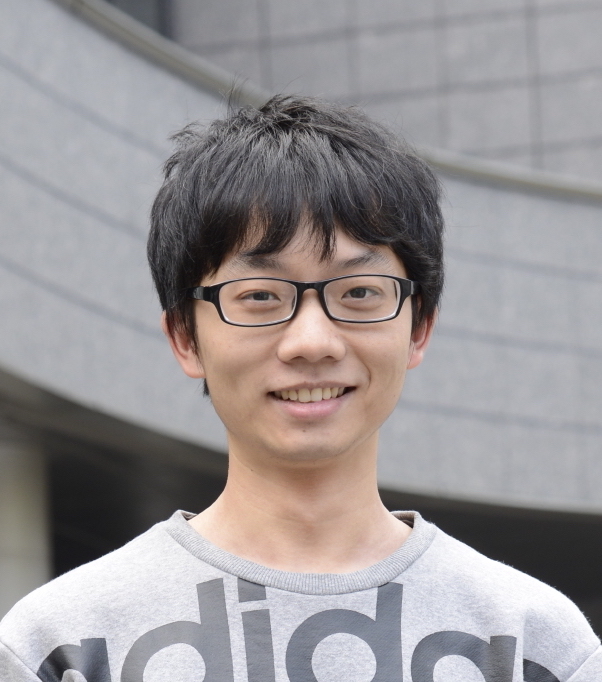}}]{Xin Wang} (S'16 - M'18)
is a project researcher at National Institute of Informatics, Japan. He received the Ph.D. degree from SOKENDAI, Japan, in 2018. Before that, he received M.S. and B.E degrees from University of Science and Technology of China and University of Electronic Science and Technology of China in 2015 and 2012, respectively. His research interests include statistical speech synthesis and machine learning.
\end{IEEEbiography}
\begin{IEEEbiography}[{\includegraphics[width=1in,height=1.25in,clip,keepaspectratio]{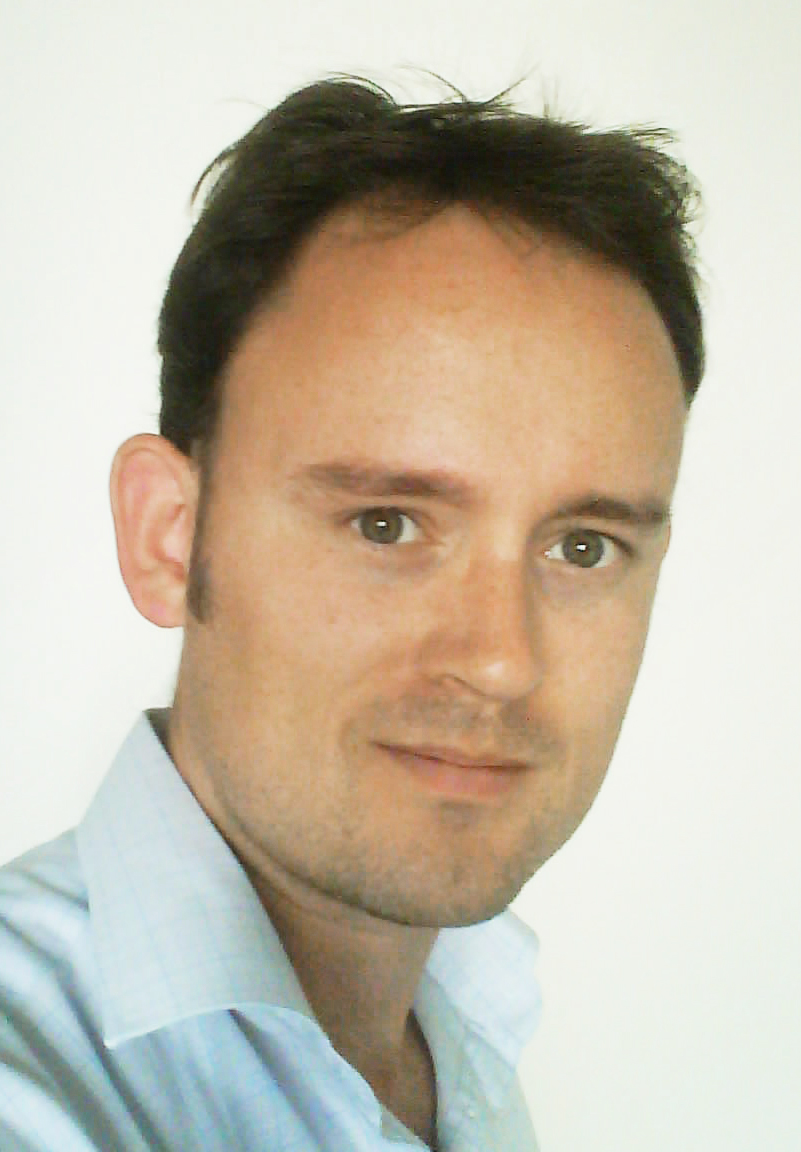}}]{Nicholas Evans} is a Professor at EURECOM, France, where he heads research in Audio Security and Privacy. He is a co-founder of the community-led, ASVspoof Challenge series and has lead or co-lead a number of special issues and sessions with an anti-spooing theme. He participated in the EU FP7 Tabula Rasa and EU H2020 OCTAVE projects, both involving anti-spoofing. Today, his team is leading the EU H2020 TReSPAsS-ETN project, a training initiative in security and privacy for multiple biometric traits.  He co-edited the second edition of the Handbook of Biometric Anti-Spoofing, served previously on the IEEE Speech and Language Technical Committee and serves currently as an asscociate editor for the IEEE Trans.\ on Biometrics, Behavior, and Identity Science.
\end{IEEEbiography}
\begin{IEEEbiography}[{\includegraphics[width=1in,height=1.25in,clip,keepaspectratio]{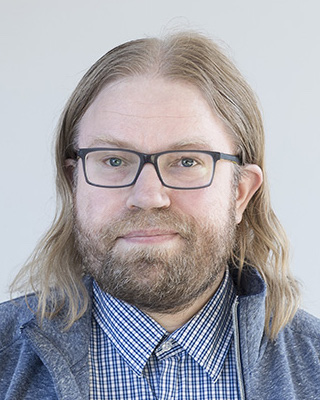}}]{Tomi H. Kinnunen} is an Associate Professor at the University of Eastern Finland. He received his Ph.D. degree in computer science from the University of Joensuu in 2005. From 2005 to 2007, he was an Associate Scientist at the Institute for Infocomm Research (I2R), Singapore. Since 2007, he has been with UEF. From 2010-2012, he was funded by a postdoctoral grant from the Academy of Finland. He has been a PI or co-PI in three other large Academy of Finland-funded projects and a partner in the H2020-funded OCTAVE project. He chaired the \emph{Odyssey} workshop in 2014. From 2015 to 2018, he served as an Associate Editor for IEEE/ACM Trans. on Audio, Speech and Language Processing and from 2016 to 2018 as a Subject Editor in \emph{Speech Communication}. In 2015 and 2016, he visited the National Institute of Informatics, Japan, for 6 months under a mobility grant from the Academy of Finland, with a focus on voice conversion and spoofing. Since 2017, he has been Associate Professor at UEF, where he leads the Computational Speech Group. He is one of the cofounders of the ASVspoof challenge, a nonprofit initiative that seeks to evaluate and improve the security of voice biometric solutions under spoofing attacks.
\end{IEEEbiography}
\begin{IEEEbiography}[{\includegraphics[width=1in,height=1.25in,clip,keepaspectratio]{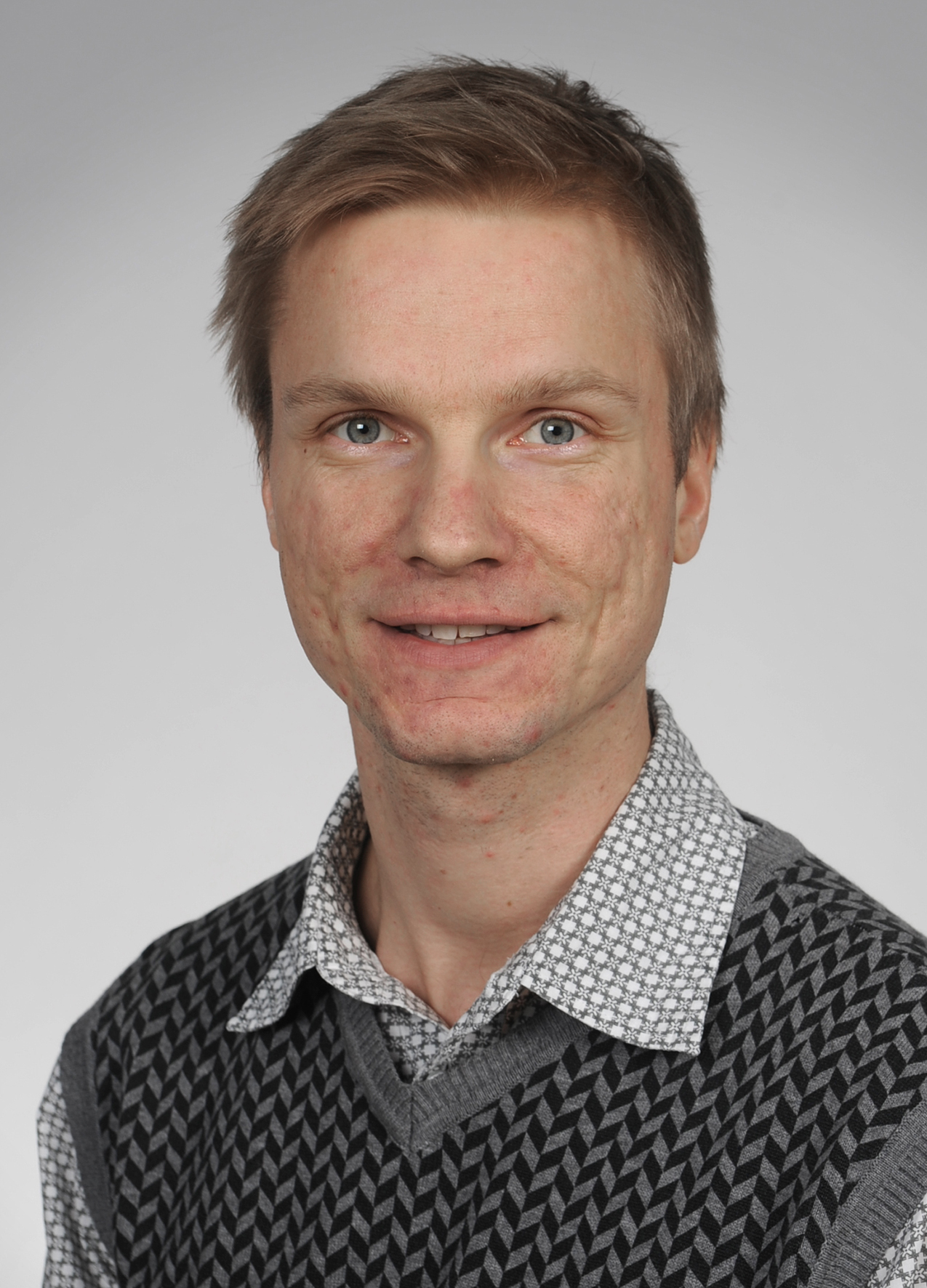}}]{Ville Vestman} is an Early Stage Researcher at the University of Eastern Finland (UEF). He received his M.S. degree in mathematics from UEF in 2013. Since 2015, his research work at UEF has been focused on speech technology and, more specifically, on speaker recognition. He is one of the co-organizers of the ASVspoof 2019 challenge.
\end{IEEEbiography}
\begin{IEEEbiography}[{\includegraphics[width=1in,height=1.25in,clip,keepaspectratio]{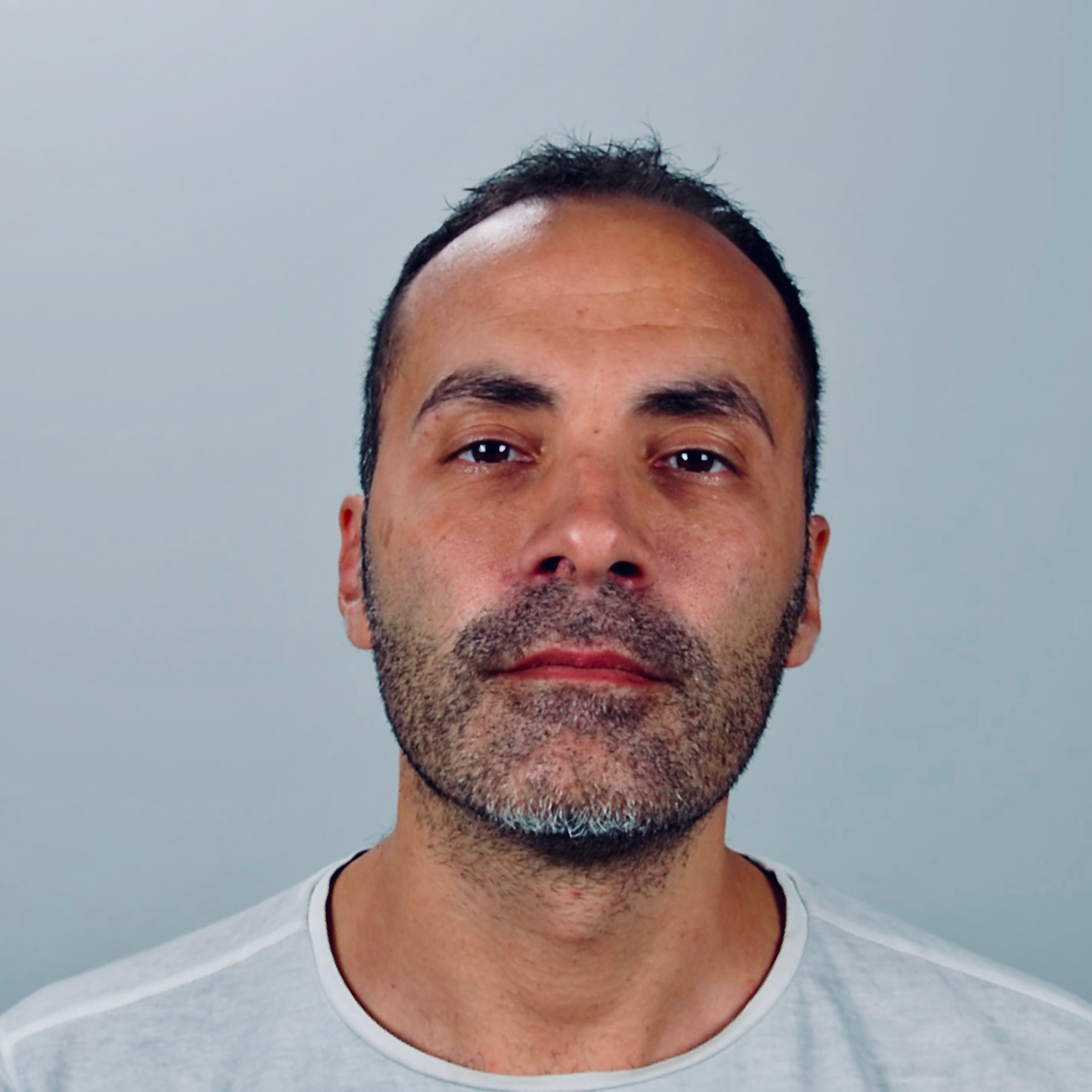}}]{Massimiliano Todisco} is an Assistant Professor within the Digital Security Department at EURECOM, France. He received his Ph.D. degree in Sensorial and Learning Systems Engineering from the University of Rome Tor Vergata in 2012. Currently, he is serving as principal investigator and coordinator for TReSPAsS-ETN, a H2020 Marie Skłodowska-Curie Innovative Training Network (ITN) and RESPECT, a PRCI project funded by the French ANR and the German DFG. He co-organises the ASVspoof challenge series, which is community-led challenges which promote the development of countermeasures to protect automatic speaker verification (ASV) from the threat of spoofing. He is the inventor of constant Q cepstral coefficients (CQCC), the most commonly used anti-spoofing features for speaker verification and first author of the highest-cited technical contribution in the field in the last three years. He has more than 90 publications. His current interests are in developing end-to-end architectures for speech processing and speaker recognition, fake audio detection and anti-spoofing, and the development of privacy preservation algorithms for speech signals based on encryption solutions that support computation upon signals, templates and models in the encrypted domain.\end{IEEEbiography}
\begin{IEEEbiography}[{\includegraphics[width=1in,height=1.25in,clip,keepaspectratio]{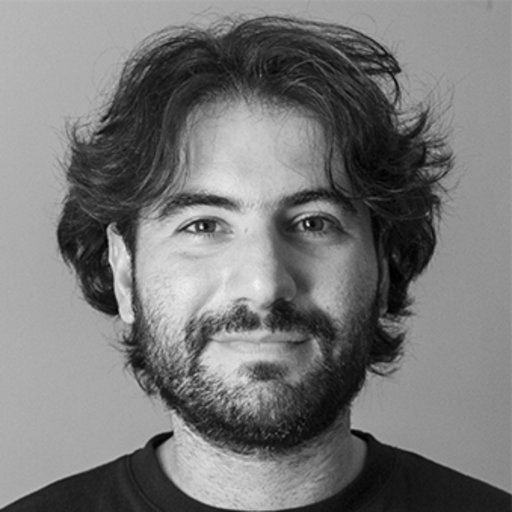}}]{H\'ector Delgado} received his Ph.D. degree in Telecommunication and System Engineering from the Autonomous University of Barcelona (UAB), Spain, in 2015. From 2015 to 2019 he was with the Speech and Audio Processing Research Group at EURECOM (France). Since 2019 he is a Senior Research Scientist at Nuance Communications Inc. He serves as an associate editor for the EURASIP Journal on Audio, Speech, and Music Processing. He is a co-organiser of the ASVspoof challenge since its 2017 edition. His research interests include signal processing and machine learning applied to speaker recognition and
diarization, speaker recognition anti-spoofing and audio segmentation.
\end{IEEEbiography}
\begin{IEEEbiography}[{\includegraphics[width=1in,height=1.25in,clip,keepaspectratio]{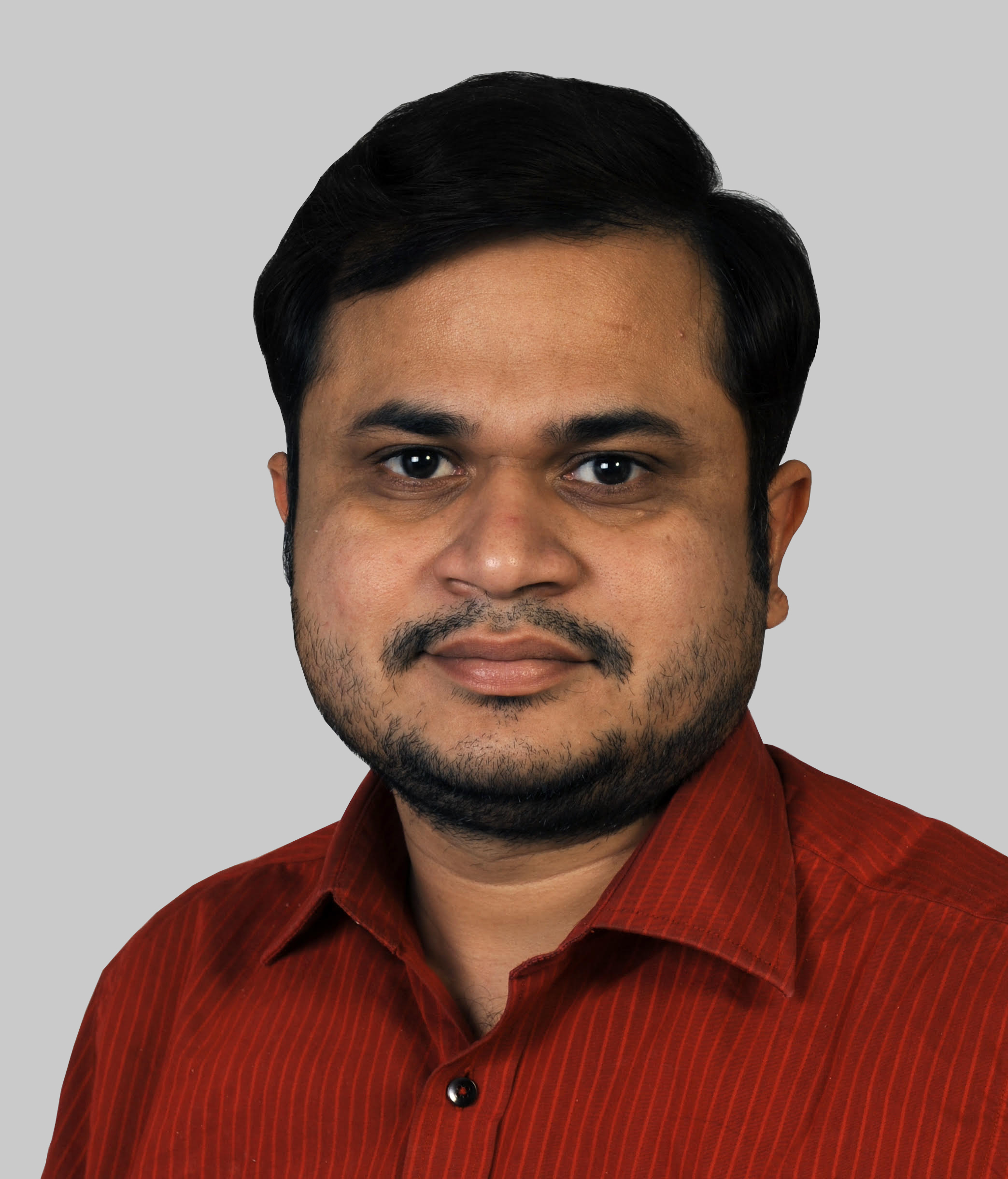}}]{Md Sahidullah} (S’09, M'15) received his Ph.D. degree in the area of speech processing from the Department of Electronics \& Electrical Communication Engineering, Indian Institute of Technology Kharagpur in 2015. Prior to that he obtained the Bachelors of Engineering degree in Electronics and Communication Engineering from Vidyasagar University in 2004 and the Masters of Engineering degree in Computer Science and Engineering from West Bengal University of Technology in 2006. In 2014-2017, he was a postdoctoral researcher with the School of Computing, University of Eastern Finland. In January 2018, he joined MULTISPEECH team, Inria, France as a post-doctoral researcher where he currently holds a starting research position. His research interest includes robust speaker recognition and spoofing countermeasures. He is also part of the organizing team of two Automatic Speaker Verification Spoofing and Countermeasures Challenges: ASVspoof 2017 and ASVspoof 2019. Presently, he is also serving as Associate Editor for the IET Signal Processing and Circuits, Systems, and Signal Processing.
\end{IEEEbiography}
\begin{IEEEbiography}[{\includegraphics[width=1in,height=1.25in,clip,keepaspectratio]{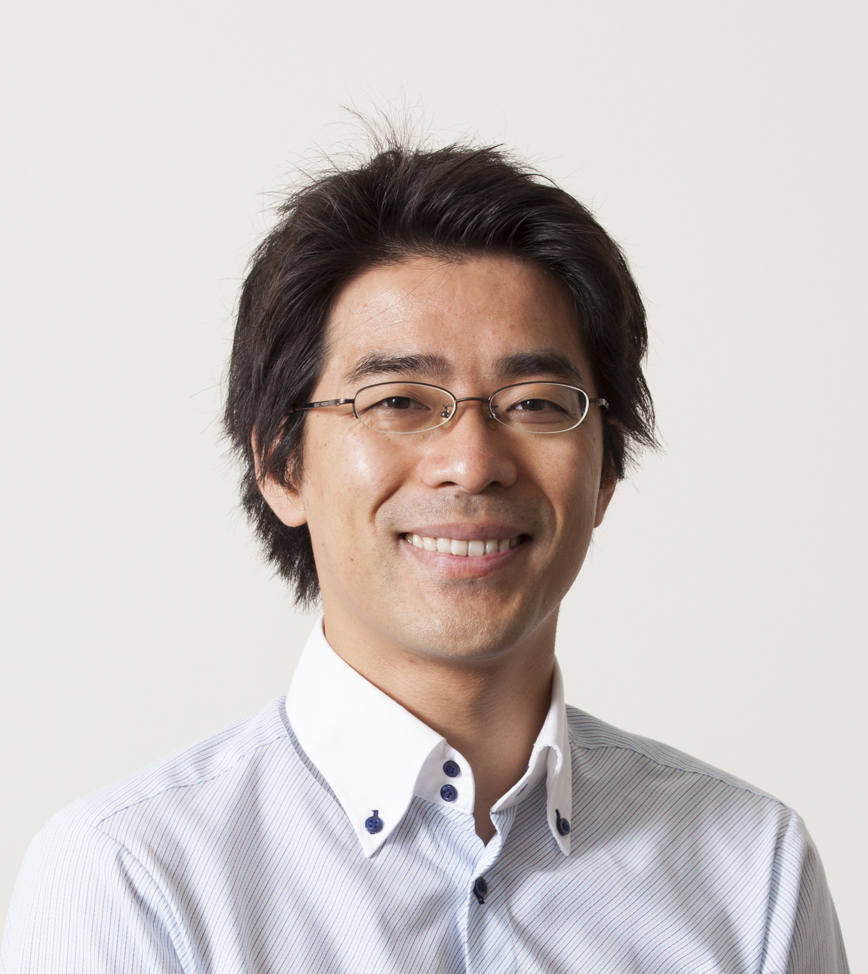}}]{Junichi Yamagishi} (SM'13) is a professor at National Institute of Informatics in Japan. He is also a senior research fellow in the Centre for Speech Technology Research (CSTR) at the University of Edinburgh, UK. He was awarded a Ph.D.\ by Tokyo Institute of Technology in 2006 for a thesis that pioneered speaker-adaptive speech synthesis and was awarded the Tejima Prize as the best Ph.D.\ thesis of Tokyo Institute of Technology in 2007. Since 2006, he has authored and co-authored over 250 refereed papers in international journals and conferences. He was awarded the Itakura Prize from the Acoustic Society of Japan, the Kiyasu Special Industrial Achievement Award from the Information Processing Society of Japan, and the Young Scientists’ Prize from the Minister of Education, Science and Technology, the JSPS prize, the Docomo mobile science award in 2010, 2013, 2014, 2016, and 2018, respectively. He served previously as co-organizer for the bi-annual ASVspoof special sessions at INTERSPEECH 2013-9, the bi-annual Voice conversion challenge at INTERSPEECH 2016 and Odyssey 2018, an organizing committee member for the 10th ISCA Speech Synthesis Workshop 2019 and a technical program committee member for IEEE ASRU 2019. He also served as a member of the IEEE Speech and Language Technical Committee, as an Associate Editor of the IEEE/ACM TASLP and a Lead Guest Editor for the IEEE JSTSP SI on Spoofing and Countermeasures for Automatic Speaker Verification. He is currently a PI of JST-CREST and ANR supported VoicePersonae project. He also serves as a chairperson of ISCA SynSIG and as a Senior Area Editor of the IEEE/ACM TASLP.
\end{IEEEbiography}
\begin{IEEEbiography}
[{\includegraphics[width=1.2in,height=1.0in,clip,keepaspectratio]{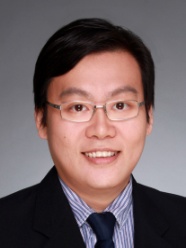}}]
{Kong Aik Lee} (M'05-SM'16) is currently a Senior Scientist at Institute for Infocomm Research, A*STAR, Singapore. From 2018 to 2020, he was a Senior Principal Researcher at the Biometrics Research Laboratories, NEC Corp., Japan. He received his Ph.D. degree from Nanyang Technological University, Singapore, in 2006. From 2006 to 2018, he was a Scientist at the Human Language Technology department, I$^2$R, A*STAR, Singapore, where he led the speaker recognition group. He was the recipient of Singapore IES Prestigious Engineering Achievement Award 2013 for his contribution to voice biometrics technology, and the Outstanding Service Award by IEEE ICME 2020. He was the Lead Guest Editor for the CSL SI on ``Two decades into Speaker Recognition Evaluation - are we there yet?'' Currently, he serves as an Editorial Board Member for Elsevier Computer Speech and Language (2016 - present), and an Associate Editor for IEEE/ACM Transactions on Audio, Speech and Language Processing (2017 - present). He is an elected member of IEEE Speech and Language Technical Committee, and the General Chair of the Speaker Odyssey 2020 Workshop.
\end{IEEEbiography}

\end{document}